\documentclass[fleqn,usenatbib]{mnras}

\usepackage{mathptmx}
\usepackage[T1]{fontenc}
\usepackage{ae,aecompl}
\usepackage{graphicx}   % Including figure files
\usepackage{amsmath}    % Advanced maths commands
\usepackage{amssymb}    % Extra maths symbols
\usepackage{verbatim} % Comment environment
\usepackage[english]{babel}
\usepackage[dvipsnames]{xcolor}
\usepackage{natbib}
\usepackage{txfonts}
\usepackage{pifont}
\usepackage{longtable} % Allows table to span multiple pages
\usepackage{booktabs} %jay

\title[Photometric, kinematic and variability study in the young open cluster NGC\,1960]
      {Photometric, kinematic and variability study in the young open cluster NGC\,1960}
\author[Y.~C.~Joshi et al.]{Y.~C.~Joshi$^{1}$\thanks{E-mail: yogesh@aries.res.in},
J. Maurya$^{1,2}$,
Ancy A. John$^{1,3}$,
A. Panchal$^{1}$,
Santosh Joshi$^{1}$,
Brijesh Kumar$^{1}$
\\
$^{1}$Aryabhatta Research Institute of Observational Sciences (ARIES), Manora peak, Nainital - 263002, India     \\
$^{2}$School of Studies in Physics and Astrophysics, Pandit Ravishankar Shukla University, Chattisgarh 492 010, India  \\
$^{3}$Pondicherry University, R. V. Nagar, Kalapet, Puducherry - 605014, India
}
\begin{document}

\date{Accepted 2019 December 20. Received 2019 December 19; in original form 2019 December 2}

\pagerange{\pageref{firstpage}--\pageref{lastpage}} \pubyear{2019}

\maketitle

\label{firstpage}

\begin{abstract}
We present a comprehensive photometric analysis of a young open cluster NGC\,1960 (= M36) along with the long-term variability study of this cluster. Based on the kinematic data of Gaia DR2, the membership probabilities of 3871 stars are ascertained in the cluster field among which 262 stars are found to be cluster members. Considering the kinematic and trigonometric measurements of the cluster members, we estimate a mean cluster parallax of 0.86$\pm$0.05 mas and mean proper motions of $\mu_{RA}$ = -0.143$\pm$0.008 mas~yr$^{-1}$, $\mu_{Dec}$ = -3.395$\pm$0.008 mas~yr$^{-1}$. We obtain basic parameters of the cluster such as $E(B-V)$ = 0.24$\pm$0.02 mag, log(Age/yr)=7.44$\pm$0.02, and $d$ = 1.17$\pm$0.06 kpc. The mass function slope in the cluster for the stars in the mass range of 0.72-7.32~$M_{\odot}$ is found to be $\gamma$ = -1.26$\pm$0.19. We find that mass segregation is still taking place in the cluster which is yet to be dynamically relaxed. This work also presents first high-precision variability survey in the central  $13^{\prime}\times13^{\prime}$ region of the cluster. The $V$ band photometric data accumulated on 43 nights over a period of more than 3 years reveals 76 variable stars among which 72 are periodic variables. Among them, 59 are short-period ($P<1$ day)and 13 are long-period ($P>1$ day). The variable stars have $V$ magnitudes ranging between 9.1 to 19.4 mag and periods between 41 minutes to 10.74 days. On the basis of their locations in the H-R diagram, periods and characteristic light curves, the 20 periodic variables belong to the cluster. We classified them as 2 $\delta$-Scuti, 3 $\gamma$-Dor, 2 slowly pulsating B stars, 5 rotational variables, 2 non-pulsating B stars and 6 as miscellaneous variables.
\end{abstract}
\section{INTRODUCTION} \label{intro}
Star clusters are important building blocks of the galaxies and it is widely believed that majority of stars in our Galaxy are formed in the star clusters. Hence study of Galactic open clusters is important for understanding the history of star formation and nature of the parent star clusters. The parameters such as age, distance, reddening, and metallicity in addition to stellar models are key to understand star formation history while luminosity and mass functions are important quantities to know their dynamical evolution \citep{1986MNRAS.220..383S, 2003ARA&A..41...57L}. The observations of large number of open clusters having different ages, locations, and environments in the Galaxy have been used to probe the Galactic structure \citep{1998MNRAS.296.1045C, 2003AJ....125.1397C, 2005MNRAS.362.1259J, 2006A&A...445..545P, 2007MNRAS.378..768J, 2019MNRAS.tmp.1972P}. While the photometric and kinematic studies of young open clusters provide clues to the star formation processes \citep{2019ApJ...870...32K}, old open clusters furnish details about the past history of the Galaxy \citep{1994ApJS...90...31P, 2016A&A...593A.116J}. These information contribute to constrain the Galaxy formation models and chemodynamical properties of the Galactic disk \citep{2008AJ....136..118F, 2018MNRAS.481.5350S, 2019AJ....158...35S}.

Since most of the open clusters are primarily affected by the field star contamination, the knowledge of membership of the stars in the cluster field is absolutely necessary to investigate the cluster properties. However, this is not the case for majority of the open clusters \citep[e.g.,][]{2002A&A...389..871D, 2008MNRAS.386.1625C, 2013A&A...558A..53K, 2016A&A...593A.116J}. Therefore, a long-term observational program is being carried out at ARIES, Nainital to better characterize some poorly studied clusters, particularly young and intermediate age open clusters [log(t/yr)<9], and determine their basic astrophysical parameters \citep{2012MNRAS.419.2379J, 2014MNRAS.437..804J}. Higher priority has been given to those open clusters for which no variability study has been carried out until now so that we can also characterize variable stars in these clusters. In this paper, we aim to determine physical parameters of one relatively young open cluster NGC\,1960 located in the Galactic anti-center direction, using the ground based optical observations supported by the archival data. Some basic parameters of the cluster are summarized in Table~\ref{webda}. This cluster has been investigated in the past in optical as well as in near-IR wavebands. The photoelectric and photographic studies of this cluster were done by \citet{1953ApJ...117..313J}, \citet{1985AZh....62..854B}, and \citet{1987A&AS...71..413M}. The proper motion study of this cluster was carried out by \citet{2000A&A...357..471S}. The photometric study of this cluster has been performed by \citet{2000A&A...357..471S, 2002A&A...383..153N, 2005A&A...438.1163K, 2006AJ....132.1669S, 2009MNRAS.399.2146W}. The near-IR photometric study of bright stars of this cluster was carried out by \citet{2008Ap&SS.313..363H}. \citet{2008AJ....135.1934S} studied the mass function and effect of photometric binaries in the cluster. Using Lithium depletion boundary technique, \citet{2013MNRAS.434.2438J} determined the age of this cluster. In spite of all these studies, many stars in the field of NGC\,1960 still lack membership confirmation that lends larger uncertainties in the estimation of cluster parameters. Recently, with the availability of the Gaia catalog \citep{2018A&A...616A...1G} having unprecedented astrometric precision, the membership determination based on kinematic method becomes a reliable tool to identify the cluster members \citep[e.g.,][]{2018A&A...618A..93C}. 

%
%Table 1: WEBDA parameters
\begin{table}
\caption{Values of parameters listed in the WEBDA for the cluster NGC~1960.}
\centering
  \label{webda}
  \begin{tabular}{cc}
  \hline
Cluster Parameters       & Values   \\ \hline
Trumpler class            & I3r      \\
RA (J2000)               & 05:36:18 \\
DEC (J2000)              & +34:08:24\\
Longitude (l/deg)        & 174.535  \\
Latitude (b/deg)         & 1.072    \\
Distance (d/pc)          & 1318     \\
Reddening (E(B-V)/mag)   & 0.222    \\
Age (Log(t/yr))          & 7.468    \\
 \hline
  \end{tabular}
\end{table}

Stars in the open clusters show different kinds of variability at various stages of their evolution with varying brightness and time scales. The photometric variability are believed to be originated through several physical mechanism like stellar pulsation, rotation of star with an inhomogeneous  distribution of cool spots, variable hot spots, obscuration by circumstellar disk, eclipsing of star, and eruption \citep[e.g.,][]{1994AJ....108.1906H, 2012MNRAS.419.2379J}. The search for variables in the open clusters is extremely important as it presents an opportunity to explore the stellar interiors. It also provides opportunity to verify stellar evolution theory and offer constraints for understanding the structure and the evolution of the Galaxy \citep{2002A&A...389..871D, 2006A&A...445..545P}. The study of variable stars in large number of open clusters have been carried out in the past; some of the recent work can be found in \citet{2012A&A...548A..97Z, 2013MNRAS.429.1466B, 2015A&A...581A..66V, 2018NewA...64...34D, 2019AJ....158...68L, 2019MNRAS.487.3505M} where large number of $\delta$-Scuti stars, $\gamma$ Doradus variables, slow pulsating B stars and other kind of variable stars are reported.  However, cluster NGC\,1960 has not been studied for variability aspect so far. As we have initiated a long-term project for the survey of variable stars in some young and intermediate age open clusters in the Galaxy in addition to accomplish their photometric study, an extensive time-series CCD observations have been carried out in the direction of the cluster NGC\,1960. Since we observed the cluster for many intra-night as well as inter-night monitoring spanning over more than three years, we probe the cluster for both short-period as well as long-period variable stars.

A detailed analysis of our photometric, kinematic and variability studies of the cluster NGC\,1960 is presented here. This paper is organized as follows: The observational and reduction techniques are presented in Sect. ~\ref{obs}. The data used in the present study is described in Sect.~\ref{data}. The kinematic study of stars in the cluster is described in Sect.~\ref{identi}. The basic parameters such as age, distance and reddening are derived in Sect.~\ref{para}. The dynamical study of the cluster is presented in Sect.~\ref{dynamical}. A detailed study of the variable stars is given in Sect.~\ref{vs} followed by their characterizations in Sect.~\ref{charac}. We discuss and summarize our results in Sect.~\ref{summary}
\section{Observations and data reduction} \label{obs}
The observations have been carried out with the 104-cm Sampurnanand Telescope (ST) at Manora Peak, Nainital. The ST is equipped with a 2k$\times$2k CCD camera having a field of view of $\sim 13^{\prime}\times13^{\prime}$ and the pixel scale of 0.758 arcsec pixel$^{-1}$ in $2\times2$ pixel binning mode. The gain of the CCD is 5.3 e$^{-}$/ADU and read out noise is 10.0 e$^{-}$. Further details of telescope and detector can be found in \citet{Joshi2005}. The bias and flat-field frames were taken on each observing night. To carry out the variability study, we monitored NGC\,1960 in $V$ band on 43 nights during the period of 2009-2013 spread over four observing seasons where we have accumulated a total of 235 frames. The exposure times range from 10 to 200 seconds depending upon the position of the target field in the sky at the time of observation, photometric sky condition, and telescope time availability. The mean PSF FWHM vary from 1.56 to 3.9 arcsec over the entire monitoring period. An observing log is given in Table~\ref{log}. The basic steps of image processing, which include bias subtraction, flat field correction, and cosmic hits removal, were performed using the standard tasks within the IRAF software. Photometry of the frames were performed using the standard {\tt DAOPHOT II} profile fitting software \citep{1992ASPC...25..297S}. To search for variable stars in the target field, absolute photometry was performed which is a meaningful tool to determine stellar parameters for the cluster members like their spectral type and stellar position in the H-R diagram. To do absolute photometry, we converted instrumental magnitudes of the stars on each night to the standard magnitudes by using the secondary standards obtained on the night of 2010, November 30 as explained in the following section. 
%
%Table 2: Observing log
\begin{table}
\vspace{-0.3cm}
\caption{The log of observations in $V$ band for the cluster NGC~1960.}
\centering
\label{log}
\begin{tabular}{cccccc}
\hline
S.N.&   Date     &   Starting JD &  $\overline{FWHM}$ & No. of & Exp. time \\
    & (yyyymmdd) &   (2450000+)  & (arcsec)  & frames &  (sec)    \\ \hline
  1 &  20091024  &  5129.398345  &   2.12    &   6    &   10      \\
  2 &  20091029  &  5134.345556  &   1.70    &   3    &   30      \\
  3 &  20091030  &  5135.258472  &   2.05    &   2    &   30      \\
  4 &  20091031  &  5136.263947  &   2.05    &   2    &   30      \\
  5 &  20091101  &  5136.457569  &   1.83    &   1    &   40      \\
  6 &  20091107  &  5143.333056  &   1.56    &   4    &   10      \\
  7 &  20091108  &  5144.413299  &   1.79    &   2    &   10      \\
  8 &  20091206  &  5172.185718  &   2.49    &   2    &   12      \\
  9 &  20100102  &  5199.351574  &   3.74    &   1    &   10      \\
 10 &  20100104  &  5201.110382  &   3.27    &   2    &   10      \\
 11 &  20100105  &  5202.088171  &   2.85    &   2    &   60      \\
 12 &  20100201  &  5229.299005  &   2.04    &   2    &   10      \\
 13 &  20100202  &  5230.282970  &   1.87    &   2    &   10      \\
 14 &  20100203  &  5231.085289  &   3.04    &   2    &   10      \\
 15 &  20100204  &  5232.247650  &   1.88    &   2    &   60      \\
 16 &  20100210  &  5234.060822  &   3.02    &   2    &   60      \\
 17 &  20100218  &  5246.126632  &   2.92    &   2    &   60      \\
 18 &  20100219  &  5247.099977  &   1.79    &   2    &   60      \\
 19 &  20100227  &  5255.199826  &   2.45    &   2    &   10      \\
 20 &  20100228  &  5256.133530  &   2.05    &   1    &   10      \\
 21 &  20100304  &  5260.190116  &   1.84    &   2    &   60      \\
 22 &  20100306  &  5262.078970  &   1.96    &   2    &   10      \\
 23 &  20100307  &  5263.152361  &   2.29    &   2    &   10      \\
 24 &  20100317  &  5273.115590  &   2.07    &   2    &   10      \\
 25 &  20100318  &  5274.099803  &   2.01    &   2    &   10      \\
 26 &  20100329  &  5285.134560  &   3.29    &   2    &   10      \\
 27 &  20100331  &  5287.107106  &   2.09    &   2    &   10      \\
 28 &  20100401  &  5288.080046  &   2.09    &   2    &   60      \\
 29 &  20101130  &  5531.347917  &   3.09    &   2    &   200     \\
 30 &  20101208  &  5539.234768  &   2.28    &   3    &   60      \\
 31 &  20110112  &  5574.217083  &   2.61    &   3    &   60      \\
 32 &  20110205  &  5598.053958  &   2.00    &   3    &   60      \\
 33 &  20110307  &  5628.158125  &   2.27    &   3    &   60     \\
 34 &  20111018  &  5853.426910  &   2.89    &   3    &   10      \\
 35 &  20111102  &  5868.445625  &   3.90    &  29    &   60      \\
 36 &  20111103  &  5869.433206  &   3.64    &  42    &   60      \\
 37 &  20111129  &  5895.234537  &   2.90    &   2    &   60      \\
 38 &  20120124  &  5951.110718  &   2.44    &  70    &   60      \\
 39 &  20120126  &  5953.124780  &   3.24    &   3    &   60      \\
 10 &  20120222  &  5980.091539  &   2.68    &   3    &   60      \\
 41 &  20120323  &  6010.113796  &   2.51    &   3    &   60      \\
 42 &  20121016  &  6218.320370  &   2.50    &   3    &   10      \\
 43 &  20130108  &  6301.075544  &   3.83    &   3    &   60      \\
\hline                                                    
\end{tabular}       
\end{table}

\section{Data sources}\label{data}
\subsection{Nainital data}\label{ntl}
To carry out detailed photometric study of the cluster NGC\,1960, we obtained Johnson-Cousins $UBVRI$ photometry of stars on 2010, November 30 using ST at Nainital. We acquired two frames each in $U$, $B$, $V$, $R$ and $I$ filters with exposure times of 300, 300, 200, 100, and 60-sec, respectively. Frames were taken when NGC\,1960 was close to zenith. We also observed two Landolt's standard fields: SA95 and PG0231+051 \citep{1992AJ....104..340L} at different airmasses on the same night. Science frames were combined together in each filter to obtain high signal-to-noise ratio (SNR) in respective filters which allowed us to obtain deeper photometries. The details of the photometric calibration along with estimation of extinction and colour coefficients are given in \citet{2012MNRAS.419.2379J, 2014MNRAS.437..804J} so we do not repeat it here. The photometric analysis of our data yields a total of 1970 stars within $\sim 13^\prime\times13^\prime$ central field of the cluster NGC\,1960. We obtained photometric data for 431, 985, 1384, 1908 and 1482 stars in the $U$, $B$, $V$, $R$ and $I$ bands, respectively. The average internal photometric errors per magnitude bin in all the five filters on the night of standardization are listed in Table~\ref{photerr}. This shows that photometric errors are relatively small ($<0.1$\,mag) for stars brighter than $V \approx 20$ mag though larger photometric errors are seen in the $U$ and $I$ bands.
\subsection{Archival data}\label{archive}
Along with our $UBVRI$ photometric catalogue, we also used many other catalogues where photometric and kinematic data were available for the cluster. To do a comprehensive study of NGC\,1960, we combined all these data along with our own data to prepare a final catalogue.
\subsubsection{Optical data}\label{phot}
Due to the limited field of view, we could observe only central $\sim 13^\prime \times 13^\prime$ region ($\sim 6^\prime.5$ radius of circular region) of NGC\,1960 while this cluster is reported to have a larger radius of about $10^\prime$ \citep{1987A&A...188...35L, 2000A&A...357..471S}, $14^\prime$ \citep{2006AJ....132.1669S} (SHA06 now onward) and $15^\prime.4$ \citep{2002A&A...383..153N}. To complement our data, particularly in the outer region of the cluster, we combined our photometric data with the SHA06 who provided $UBVRI$ photometric catalogue in the wide field of view ($\sim 50^\prime \times 50^\prime$) observed through the 105-cm Kiso Schmidt telescope, although present photometry is deeper in comparison of SHA06. Since we have taken large number of stars from SHA06, we consider their estimated cluster radius of $14^\prime$ and combined catalogue is confined to this radius only. As NGC\,1960 is a relatively nearby and younger cluster hence some of the most bright stars in the cluster field got saturated in our CCD observations as well as in SHA06. Furthermore, some stars fell into the bad CCD pixels which were rejected during the image analysis. Therefore, we acquired magnitudes of 12 such bright stars from the previous catalogues of \citet{1987A&AS...71..413M, 2000A&A...355L..27H, 2013MNRAS.434.2438J}. In this way we made a combined catalogue of 3962 stars for which photometric data has been compiled.
%
%Table 1: Mag-err
\begin{table}
\centering
\caption{The average photometric error per magnitude bin as a function of stellar brightness in Nainital data taken on 30 November 2010.}
\label{photerr}
\begin{tabular}{lccccc}
\hline
mag & $\sigma_U$ & $\sigma_B$   & $\sigma_V$  & $\sigma_R$  & $\sigma_I$\\
\hline
 8$-$ 9  &  0.006  &   -     &   -     &   -     &  0.011  \\
 9$-$10  &  0.005  &   -     &   -     &  0.009  &  0.005  \\
10$-$11  &  0.004  &  0.017  &  0.010  &  0.005  &  0.004  \\
11$-$12  &  0.005  &  0.012  &  0.005  &  0.005  &  0.005  \\
12$-$13  &  0.006  &  0.010  &  0.005  &  0.005  &  0.005  \\
13$-$14  &  0.007  &  0.010  &  0.007  &  0.005  &  0.005  \\
14$-$15  &  0.007  &  0.011  &  0.006  &  0.006  &  0.006  \\
15$-$16  &  0.010  &  0.011  &  0.007  &  0.007  &  0.009  \\
16$-$17  &  0.017  &  0.011  &  0.008  &  0.010  &  0.015  \\
17$-$18  &  0.034  &  0.014  &  0.012  &  0.019  &  0.032  \\
18$-$19  &  0.075  &  0.019  &  0.022  &  0.039  &  0.064  \\
19$-$20  &  0.160  &  0.036  &  0.047  &  0.088  &  0.159  \\
20$-$21  &  0.413  &  0.080  &  0.099  &  0.222  &   -     \\
21$-$22  &   -     &  0.171  &  0.256  &  0.514  &   -     \\
\hline
\end{tabular}
\end{table}

\subsubsection{2MASS near-IR data}\label{2mass}
We used archival near-IR photometric data from the Two Micron All-Sky Survey [$2MASS$] \citep{2006AJ....131.1163S} which provides photometry in the $J$ (1.25 $\mu$m), $H$ (1.65 $\mu$m), and $K_s$ (2.17 $\mu$m) filters. The data has limiting magnitude of 15.8, 15.1, and 14.3 mag in $J$, $H$, and $K_{s}$ bands, respectively, having a signal-to-noise ratio greater than 10. Our optical data was cross-correlated with the 2MASS photometric catalogue and found 3142 common stars within $1^{\prime\prime}$ matching radius for which we could extract $J$, $H$, and $K_s$ magnitudes. To ensure the photometric accuracy, we used only those stars having  $J$, $H$, and $K_s$ magnitudes that have quality flag ph-qual=$AAA$, which represents a SNR$\ge$10 and photometric uncertainty <0.10 mag. The $K_s$ magnitudes were converted into $K$ magnitudes using equations given in the \citet{2001AJ....121.2851C}.
\subsubsection{Gaia proper motion data}\label{Gaia}
We took data from the Gaia archive DR2 for the proper motion studies \citep{2018A&A...616A...1G}. The Gaia mission, which launched in 2013 to measure positions, trigonometric parallaxes, proper motions, and photometry of million of stars, provides mean parallax error up to 0.04 mas for sources having $G \leq 15$ mag and around 0.7 mas for sources having $G = 20$ mag. The DR2 provides proper motions of more than 1.3 billions sources with uncertainties up to 0.06 mas yr$^{-1}$ for the sources having $G \leq 15$ mag, 0.2 mas yr$^{-1}$ for G = 17 mag, and 1.2 mas yr$^{-1}$ for sources up to $G = 20$ mag. In the present analysis, we found 3871 common stars between our catalogue and Gaia DR2 catalogue within $1^{\prime\prime}$ matching radius for which we could extract the proper motions and parallax data.
%
%UBVRI photometric catalogue
% tab:photclus
%___________________________________________________________________________
\begin{table*}
\caption{Photometric catalogue of 3962 stars detected in the field of cluster NGC~1960. Table is sorted in the order of increasing $V$ magnitude. The error in magnitudes indicates the photometric error in the measurement. Column 1 gives identification number and columns 2 and 3 give right ascension and declination for J(2000). From columns 4 to 8, we provide photometric magnitudes and corresponding error in the $UBVRI$ passbands, wherever present. Full table available online further contains 2MASS $JHK$ magnitudes, Gaia parallax, proper motions and membership probability for all the stars and their associated errors.}
\label{catalog}
\centering
\begin{tabular}{cccccccc}
\hline
  ID  &     RA       &     DEC      &         U        &        B         &        V         &       R          &        I         \\
\hline
0001  &  05:36:23.05  & +34:10:32.8 &  8.276           &  9.106$\pm$0.067 &  8.291$\pm$0.122 &   -              &   -               \\
0002  &  05:36:15.79  & +34:08:36.9 &  8.219           &  8.879           &  8.880           &   -              &   -               \\
0003  &  05:36:39.24  & +34:03:50.1 &  8.482           &  9.082           &  9.060           &  -               &   -                 \\
0004  &  05:36:31.99  & +34:10:47.2 &  8.472$\pm$0.006 &  9.100           &  9.090           &   -              &  8.980$\pm$0.012  \\
0005  &  05:35:44.08  & +33:59:44.4 &   -              &  9.105$\pm$0.021 &  9.099$\pm$0.022 &   -              &   -               \\
    . &       .       &     .       &         .        &        .         &        .         &         .        &         .         \\
    . &       .       &     .       &         .        &        .         &        .         &         .        &         .         \\
    . &       .       &     .       &         .        &        .         &        .         &         .        &         .         \\
3961  &  05:35:57.98  & +34:10:41.2 &         -        &        -         &        -         & 19.630$\pm$0.078 &  18.697$\pm$0.113 \\ 
3962  &  05:36:39.29  & +34:14:17.8 &         -        &        -         &        -         & 19.947$\pm$0.128 &  18.407$\pm$0.037 \\
\hline
\end{tabular} 
\end{table*}

\subsection{Final catalogue}\label{cat}
The final photometric catalogue contains 3962 stars which comprises our $UBVRI$ data combined with the wide field photometry given by \citet{2006AJ....132.1669S} and 2MASS near-IR data \citep{2006AJ....131.1163S}. Here, we note that all the magnitudes are neither available for all the stars nor in all the passbands. To convert the pixel coordinates (X,Y) into celestial coordinates ($\alpha_{2000}$, $\delta_{2000}$), a linear astrometric solution was derived by matching common bright stars between our $V$ band frame and the Gaia DR2 catalogue. We achieved a radial RMS scatter in the residuals of $\sim$  $0^{\prime\prime}.6$, which is equivalent to $\sim$ 0.8 pixel. A sample of final catalogue is given in Table~\ref{catalog}. The entire catalogue is available online that contains star ID, celestial coordinates, photometric magnitudes in $U$, $B$, $V$, $R$, $I$ bands; 2MASS $J$, $H$, and $K$ magnitudes; the Gaia parallax ($\overline\omega$) and proper motions ($\mu_{x}$, $\mu_{y}$). Each value is given with its associated error for all the stars. The final catalogue contains stars down to $B=21.9$ and $V=21.4$ mag though photometric errors become large ($\geq 0.1)$ mag) for stars fainter than 20 mag.
\section{Membership determination}\label{identi}
The basic parameters for the cluster NGC\,1960 have been derived by several authors \citep[e.g.,][]{2000A&A...357..471S, 2002A&A...383..153N, 2006AJ....132.1669S, 2009MNRAS.399.2146W}. However, most of these authors used all the stars present in the observed field of the cluster region to determine their cluster parameters. Since not all the stars present in the region are associated with the cluster hence their parameter estimations render larger uncertainties. Therefore, in the present study, we first identified cluster members through Gaia DR2 astrometric and kinematic measurements.
%
%------------------------------------ Fig. VPD ----------------------
\begin{figure}
\centerline{\includegraphics[width=7.5cm,height=6.5cm]{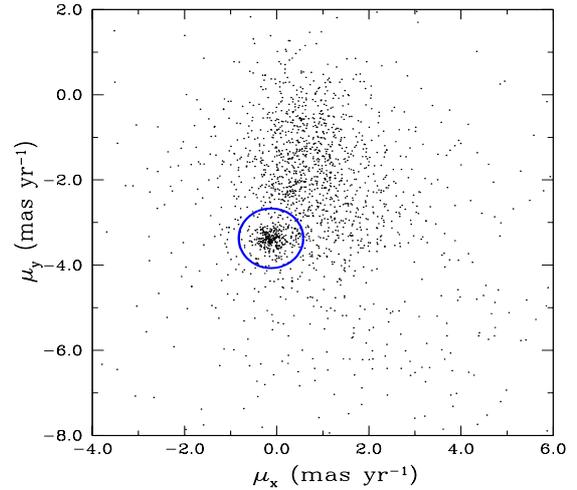}}
\caption{Vector-point diagram of the proper motions of stars constructed using Gaia DR2 proper motions in the field of the cluster NGC\,1960.}
\label{vpd}
\end{figure}
%------------------------------------ End Fig. ----------------------
%
\subsection{Membership probabilities}\label{prob}
\begin{figure}
\centerline{\includegraphics[width=8.5cm,height=6.5cm]{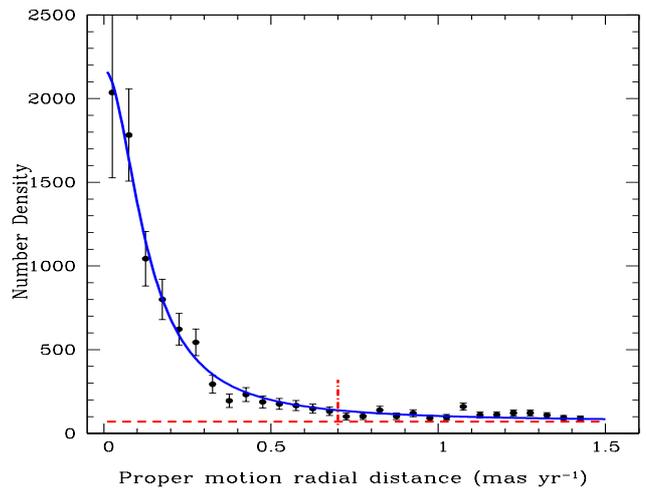}}
\caption{The radial distribution of stellar number density in the proper-motion plane. Here, dashed horizontal line indicates the field density and vertical dotted line represents the cut-off radius used to find the cluster members.}
\label{rdp_pm}
\end{figure}

In previous years many methods were used for the membership determination of stars in the star clusters based on the photometric and kinematic data \citep{2013MNRAS.430.3350Y,2014A&A...564A..79D, 2014MNRAS.437..804J, 2017MNRAS.470.3937S, 2018A&A...615A.166T}. However, availability of the astrometric data from the Gaia survey with the unprecedented accuracy has made the kinematic method of membership determination most reliable. In the present study we use proper motions from the Gaia DR2 to obtain the membership probabilities of 3962 stars found within the cluster radius. We found 3866 stars that have proper motions available in the present catalogue. Proper motions for these stars in the RA-DEC plane are plotted as Vector-point diagram (VPD) in Figure~\ref{vpd}. It is evident in the VPD that the cluster stars are well separated from the field stars. The center of the circular region confining the probable cluster members was determined by maximum density method in the proper motion plane which is found to lie at $(\mu_x, \mu_y) \equiv (\mu_\alpha cos\theta, \mu_\delta) \approx (-0.13, -3.37)$ mas~yr$^{-1}$. The radius of the circle was derived by plotting the stellar density as a function of radial distance in the proper motion plane as illustrated in Figure~\ref{rdp_pm}. We fit the stellar density profile with a function similar to the one used to characterize the radial profiles of star clusters in the galaxies. In  Figure~\ref{rdp_pm}, we draw horizontal dashed line to show stellar field density. In the radial density distribution, we put a cut-off where stellar density falls close to the field density which is found to be $\sim$ 0.7 mas yr$^{-1}$ and shown by vertical dashed line in the figure. In this way, the radius of the circle is determined as 0.7 mas yr$^{-1}$ and shown by a blue circle in Figure~\ref{vpd}. We thus obtained a total 462 stars within the circular region which could be the potential cluster members.

%------------------------------------ Fig. RDP_pm--------------------
%------------------------------------ End Fig. --------------------
%
%

To determine the membership probabilities of stars in the field of the cluster, we used a statistical method described in \citet{1998A&AS..133..387B} and \citet{2019MNRAS.487.3505M}. In this method membership probability of the i$^{th}$ star is defined as
$$P_{\mu}(i) = \frac{n_{c}~.~\phi_c^{\nu}(i)}{n_{c}~.~\phi_c^{\nu}(i) + n_f~.~\phi_f^{\nu}(i)}$$
where $n_{c}$ and $n_{f}$ are the normalized number of stars for cluster and field regions i.e. $n_c + n_f = 1$. The $\phi_c^{\nu}$ and $\phi_f^{\nu}$ are the frequency distribution functions for the cluster and field stars. The $\phi_c^{\nu}$ for the i$^{th}$ star is defined as:\\

~~~~~~~~~~~$\phi_c^{\nu}(i) =\frac{1}{2\pi\sqrt{{(\sigma_{xc}^2 + \epsilon_{xi}^2 )} {(\sigma_{yc}^2 + \epsilon_{yi}^2 )}}}$

~~~~~~~~~~~~~~~~~~~~$ exp\{{-\frac{1}{2}[\frac{(\mu_{xi} - \mu_{xc})^2}{\sigma_{xc}^2 + \epsilon_{xi}^2 } + \frac{(\mu_{yi} - \mu_{yc})^2}{\sigma_{yc}^2 + \epsilon_{yi}^2}] }\}$ \\

\noindent where $\mu_{xi}$ and $\mu_{yi}$ are the proper motions in right ascension and declination, respectively while $\epsilon_{xi}$ and $\epsilon_{yi}$ are the corresponding errors in the proper motions of $i^{th}$ star. Here, proper motion of the cluster center are $\mu_{xc}$ and $\mu_{yc}$ with dispersion $\sigma_{xc}$ and $\sigma_{yc}$. Further we define\\

$\phi_f^{\nu}(i) =\frac{1}{2\pi\sqrt{(1-\gamma^2)}\sqrt{{(\sigma_{xf}^2 + \epsilon_{xi}^2 )} {(\sigma_{yf}^2 + \epsilon_{yi}^2 )}}} exp\{{-\frac{1}{2(1-\gamma^2)}}$

$[\frac{(\mu_{xi} - \mu_{xf})^2}{\sigma_{xf}^2 + \epsilon_{xi}^2} -\frac{2\gamma(\mu_{xi} - \mu_{xf})(\mu_{yi} - \mu_{yf})} {\sqrt{(\sigma_{xf}^2 + \epsilon_{xi}^2 ) (\sigma_{yf}^2 + \epsilon_{yi}^2 )}} + \frac{(\mu_{yi} - \mu_{yf})^2}{\sigma_{yf}^2 + \epsilon_{yi}^2}]\}$\\

\noindent where $\mu_{xf}$ and  $\mu_{yf}$ are the field proper motion with dispersion $\sigma_{xf}$ and $\sigma_{yf}$. The correlation coefficient $\gamma$ is defined as\\

$\gamma = \frac{(\mu_{xi} - \mu_{xf})(\mu_{yi} - \mu_{yf})}{\sigma_{xf}\sigma_{yf}}$\\

\noindent From the VPD of the cluster, we considered 462 stars within the circle as probable cluster members and remaining 3500 stars as probable field members. We thus determined $n_{c}$ = 0.12 and $n_{f}$ = 0.88. We obtained the mean proper motions within the circular region as $\mu_{xc}$=-0.09 mas yr$^{-1}$ and $\mu_{yc}$=-3.36 mas yr$^{-1}$ with corresponding dispersion $\sigma_{xc}$ = 0.28 mas yr$^{-1}$ and $\sigma_{yc}$ = 0.26 mas yr$^{-1}$. The mean proper motions of the probable field stars were found as $\mu_{xf}$=0.94 mas yr$^{-1}$ and $\mu_{yf}$=-2.55 mas yr$^{-1}$ with corresponding dispersion $\sigma_{xf}$ = 2.86 mas yr$^{-1}$  and $\sigma_{yf}$ = 3.58 mas yr$^{-1}$. Using the above formulae, we estimated membership probabilities of all the stars lying within the cluster region except 91 stars which have no proper motion information available in the Gaia DR2 data.
%
% List of cluster members
  \begin{table*}
    \caption{The photometric parameters given for 262 cluster members. Membership probability estimated through kinematic study is given in the last column.}
    \label{mpm}
    \small
    \begin{center}
\begin{tabular}{rcccccccc}
\hline
$ID$   &     RA         &      DEC        &   $V$              & $(B-V)$         &($\overline\omega$) &     $\mu_{RA}$     &    $\mu_{Dec}$     &  Prob  \\
       &   (J2000)      &    (J2000)      &   (mag)            & (mag)           &       (mas)        &  (mas yr$^{-1}$)   &   (mas yr$^{-1}$)  &        \\ \hline
0003   &   05:36:39.24  &  +34:03:50.1    &   9.060            &  0.022          &   0.789$\pm$0.062  &  0.104$\pm$0.107   &  -3.864$\pm$0.075  &   0.74 \\
0004   &   05:36:31.99  &  +34:10:47.2    &   9.090            &  0.010          &   0.756$\pm$0.050  &  0.299$\pm$0.100   &  -3.491$\pm$0.075  &   0.87 \\
0006   &   05:36:22.59  &  +34:08:02.0    &   9.150            & -0.002          &   0.782$\pm$0.065  & -0.151$\pm$0.131   &  -3.451$\pm$0.097  &   0.94 \\
0007   &   05:36:42.30  &  +34:12:06.0    &   9.250            &  0.050          &   0.764$\pm$0.058  &  0.159$\pm$0.107   &  -3.519$\pm$0.081  &   0.92 \\
0009   &   05:36:21.94  &  +34:08:08.2    &   9.374$\pm$0.031  &  0.008$\pm$0.042&   0.940$\pm$0.064  & -0.134$\pm$0.128   &  -3.504$\pm$0.097  &   0.94 \\
 .     &       .        &      .          &    .    .          &   .             &    .    .          &   .    .           &    .    .          &    .   \\
 .     &       .        &      .          &    .    .          &   .             &    .    .          &   .    .           &    .    .          &    .   \\
 .     &       .        &      .          &    .    .          &   .             &    .    .          &   .    .           &    .    .          &    .   \\
3548   &   05:35:38.97  &  +34:12:34.6    &  20.176$\pm$0.140  &   -             &   0.829$\pm$0.470  &  0.192$\pm$0.824   &  -3.220$\pm$0.711  &   0.68 \\
3902   &   05:36:46.69  &  +34:16:58.5    &  20.650$\pm$0.337  &   -             &   0.702$\pm$0.379  & -0.524$\pm$1.014   &  -3.445$\pm$0.712  &   0.66 \\
\hline
\end{tabular}
    \end{center}
  \end{table*}
%  \end{landscape}
%----------------------------------------- Table 4 ----

%
\subsection{Parallax criteria on membership selection}
\begin{figure}
\centerline{\includegraphics[width=8.5cm,height=6.5cm]{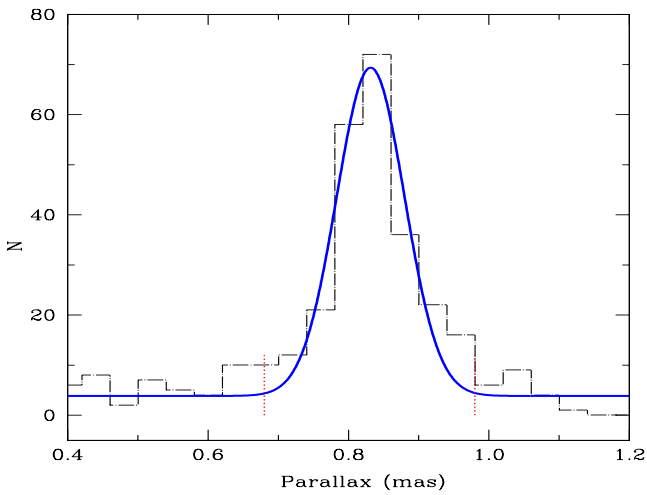}}
\caption{The parallax distribution of 462 stars shown in encircled region in Figure~\ref{vpd}. The bin size is taken as 0.04 mas. The thick line represents the best fit Gaussian profile. The two vertical dotted lines exhibit $3\sigma$ cut-off limits around the peak value.}
\label{parallax}
\end{figure}

An additional check on our selection of cluster members is done through parallax measurements ($\overline\omega$) provided by the Gaia DR2 catalogue. In Figure~\ref{parallax}, we illustrate histogram of the parallax measurements of these 462 probable members. Here, we used only those stars for which error in parallax was smaller than 0.2 mas. The mean value of the parallax is derived by fitting a Gaussian profile on the histogram shown by a continuous line in the figure. The peak and standard deviation $\sigma$ of the parallax distribution are found to be 0.83 mas and 0.05 mas, respectively as estimated from the best fit Gaussian profile.

Recently, \citet{2018A&A...616A...2L} reported a general offset in Gaia parallaxes by -0.029 mas though there is also some evidence that the offset increases for the distances larger than 1 kpc \citep{2018ApJ...862...61S, 2018MNRAS.478.3825L}. This has been further confirmed by many other surveys although with slightly different values \citep[e.g.,][]{2019MNRAS.487.3568S, 2019ApJ...878..136Z}. Applying this offset to our estimate of mean parallax, we found a mean parallax of 0.86$\pm$0.05 mas which corresponds a distance of $\sim$ 1.17$\pm$0.06 kpc [$(m-M)_0 = 10.33\pm0.11$ mag] for the cluster. Our estimate from Gaia DR2 thus suggests a slightly smaller distance for NGC\,1960 than the distance of 1.32$\pm$0.12 kpc obtained by \citet{2000A&A...357..471S}, 1.31 kpc obtained by \citet{2005A&A...438.1163K} and 1.33 kpc obtained by \citet{2006AJ....132.1669S, 2009MNRAS.399.2146W}.

%------------------------------ Fig Parallax ------------------------
%------------------------------------------------------------------------------
% 

To further isolate cluster members from the contamination of field stars, we used mean cluster parallax as a second check. We eliminated all those stars which deviate from the mean parallax by more than 3$\sigma$. We thus found 263 stars out of 462 stars which lie within this region. Interestingly, we found membership probabilities of all these stars above 60\% except three stars for which membership probability lies in between 48 to 52\%. As in some previous studies  \citep[e.g.,][]{2017AcA....67..203R} it was suggested that even if membership probability based on proper motions is slightly smaller but star has higher geometric probability (position with respect to cluster center) and photometric probability (location in the colour-magnitude diagram), the star could still be a cluster member. We therefore further examined these 3 stars on the basis of their spatial positions, locations in the $(B-V)/V$ and $(V-I)/V$ colour-magnitude diagrams and $(U-B)/(B-V)$ colour-colour diagram. Two of these three stars are found to be good candidates for the cluster members. Therefore, we considered 262 stars as the cluster members which are used in the subsequent analysis.
\subsection{Mean proper motions of the cluster}
\begin{figure}
\centerline{\includegraphics[width=8.5cm,height=10.0cm]{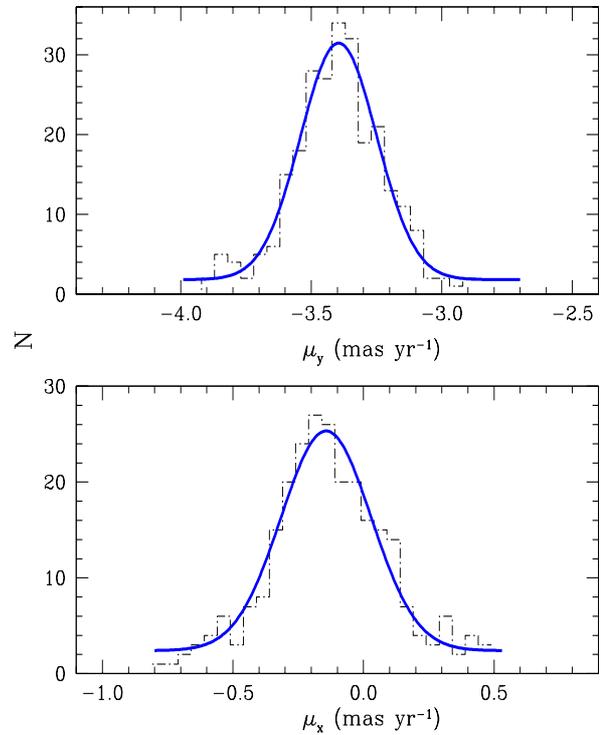}}
\caption{Proper motion histograms with a bin size of 0.05 mas yr$^{-1}$ for the 262 cluster members. The Gaussian fit shown by thick continuous lines are drawn to determine mean proper motions of the cluster NGC\,1960.}
\label{pm_hist}
\end{figure}

To estimate the mean proper motions, we draw an histogram of proper motions of 262 cluster members in the x- and y- directions in Figure~\ref{pm_hist}. We fit a Gaussian profile in the distributions and mean value of the proper motion is estimated corresponding to the peak in the distribution. The mean proper motions in right ascension ($\bar{\mu_{x}}$) and declination ($\bar{\mu_{y}}$), respectively are found to be
$$
\bar{\mu_{x}}=-0.143\pm0.008~mas/yr,~~~~ \bar{\mu_{y}}=-3.395\pm0.008~mas/yr
$$
The mean proper motion of the cluster is determined as $(\bar\mu_{x}^2+\bar\mu_{y}^2)^{1/2}$ which is found to be 3.398$\pm$0.011 mas yr$^{-1}$. From the radial-velocity measurement of 114 stars computed from the Tycho-2 catalogue, \citet{2003ARep...47....6L} estimated a proper motions of $\bar{\mu_{x}}=0.99\pm0.17$~mas/yr and $\bar{\mu_{y}}=-3.96\pm0.15$~mas/yr for the cluster. \citet{2005A&A...438.1163K} have also determined the mean proper motions for this cluster as $0.50$~mas/yr and $-4.94$~mas/yr in $RA$ and $DEC$ directions, respectively. It is thus found that previous reported values in the literature were slightly overestimated.

%------------------------------------ Fig. PM_hist--------------------
%------------------------------------ End Fig. 03 --------------------
 
The sample of cluster members in NGC\,1960 is given in Table~\ref{mpm} which provides star ID, magnitude, colour, parallax, mean proper motions and membership probabilities for these stars and full catalogue is available online.
\section{Basic parameters of NGC\,1960}\label{para}
\subsection{Extinction measurement}\label{ext}
\subsubsection{Reddening in optical bands}\label{red_opt}
\begin{figure}
\centerline{\includegraphics[width=8.5cm,height=6.5cm]{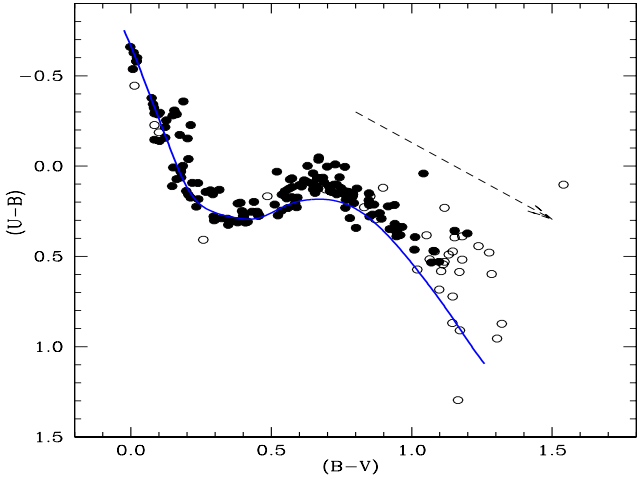}}
\caption{The $(U-B, B-V)$ colour-colour diagram for the cluster members for which simultaneous $UBV$ magnitudes are available. Here, open circles represent the stars having photometric errors larger than 0.05 mag in $U$, 0.015 mag in $B$ and 0.01 mag in $V$ bands. The dashed arrow represents the slope ($0.84\pm0.02$) and direction of the reddening vector. The solid line represents the Zero-Age Main Sequence taken from \citet{1982ND...26..14S}.}
\label{bv_ub}
\end{figure}

The reddening, $E(B-V)$, in the field of cluster NGC\,1960 can be estimated using the $(U-B)$ vs $(B-V)$  two-colour diagram (TCD). In our list of 262 cluster members, we found only 185 stars for which simultaneous $U$, $B$ and $V$ magnitudes are available. In Figure~\ref{bv_ub}, we illustrate $(U-B)/(B-V)$ diagram for these stars. Here, we also draw \citet{1982ND...26..14S} zero-age-main-sequence in the figure by a solid line. Lacking any prior estimate of the metallicity in the literature, we conservatively adopted a solar metallicity for the cluster. To determine the reddening, we primarily focus on those stars which have spectral type earlier than $A0$ as later type stars may be more affected by the metallicity and background contamination \citep{2003MNRAS.345..269H}. We found a reddening vector of $E(U-B)/E(B-V)=0.84\pm0.02$ in the direction of this cluster which is slightly larger than the standard reddening law of $E(U-B)/E(B-V)=0.72$ given by \citet{1953ApJ...117..313J}. From the visual best fit in the $(U-B)$ vs $(B-V)$ diagram, we estimated a reddening $E(B-V)=0.24\pm0.02$ mag. The photometric study of NGC\,1960 by \citet{2006AJ....132.1669S} suggested a non-uniform reddening across the cluster region which is apparent in the distribution of stars in the $(U-B)/(B-V)$ diagram. Our reddening estimate is consistent with the 0.22 mag obtained by \citet{2006AJ....132.1669S} and $0.25\pm0.02$ given by \citet{2000A&A...357..471S} but larger than $0.20\pm0.02$ mag determined by \citet{2008MNRAS.386..261M}. Assuming a standard reddening law, the colour-excess $E(V-I)$ was estimated as $0.30\pm0.02$ mag using the relation $E(V-I)=1.25~\times E(B-V)$ \citep{1989ApJ...345..245C}.
%
%-----------------------------Fig. BV_UB ------------------------
%------------------------------------------------------------------------------
%
\subsubsection{Reddening in near-IR bands}\label{red_nir}
\begin{figure}
\centerline{\includegraphics[width=8.5cm,height=6.5cm]{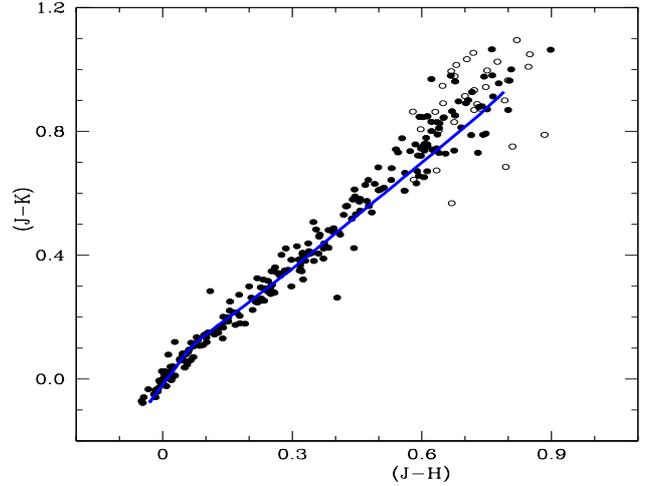}}
\caption{The $(J-H)/(J-K)$ diagram for cluster members. Here, open circles represent the stars which have photometric error larger than 0.05 mag in $J$, 0.15 mag in $H$ and 0.15 mag in $K$ bands. The solid line represents the best fit solar metallicity isochrones.}
\label{infra_tcd}
\end{figure}

Since NGC\,1960 is a young open cluster and possibly still embedded in the parent molecular cloud, it would be more appropriate if we determine reddening in only near-IR bands. We therefore draw the $(J-H)/(J-K)$ colour-colour diagram which is shown in Figure~\ref{infra_tcd}. We also overplot Marigo isochrones of solar metallicity \citep{2017ApJ...835...77M} by shifting the line in the direction of reddening vector. A best fit was achieved by shifting $E(J-H)$ = 0.07 mag with the ratio $\frac{E(J-H)}{E(J-K)}$ = 0.65 for the cluster. The reddening vector derived in the present study is slightly higher than the usual interstellar extinction ratio of 0.55 given by \citet{1989ApJ...345..245C}. $E(B-V)$ can be estimated from the near-IR reddening using the following relation:
$$E(J-H) = 0.309 \times E(B-V)$$
The reddening $E(B-V)$ was determined as 0.23 mag for the cluster NGC\,1960. It is thus found that the reddening $E(B-V)$ derived from the near-IR photometry is in excellent agreement with the value derived from the optical photometry.
Although there is some evidence of non-uniform extinction present in this cluster from the optical TCD but a consistent reddening measurement between the optical and near-IR bands suggests that non-uniform or differential extinction is not significant within the cluster.
%
%---------------------------- Fig. Infrared_TCD--------------------
%------------------------------------ End Fig. --------------------
% 
\subsection{Colour-magnitude diagrams and age determination}\label{cmd}
\begin{figure}
\centerline{\includegraphics[width=8.5cm,height=5.5cm]{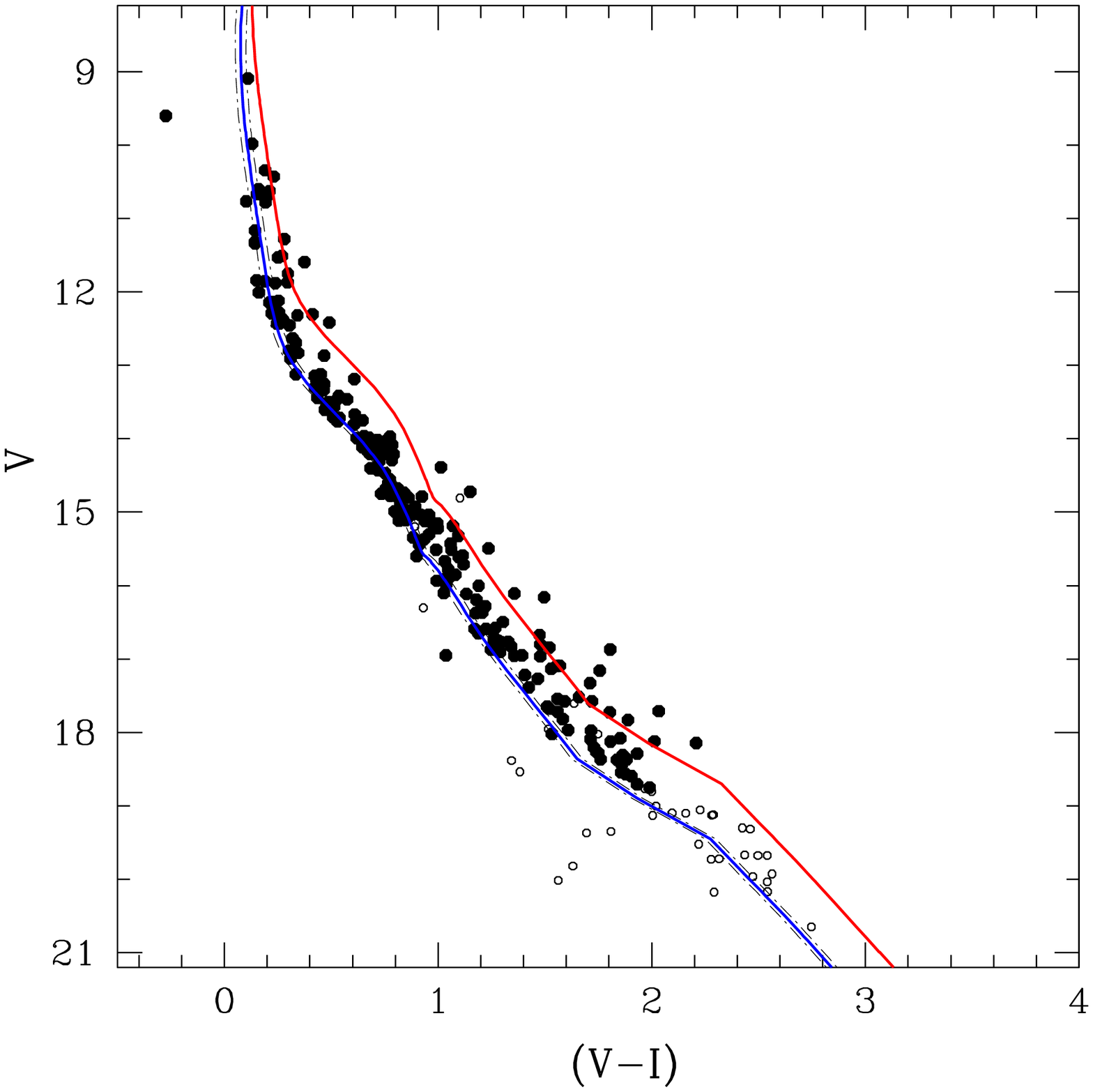}}
\centerline{\includegraphics[width=8.5cm,height=5.5cm]{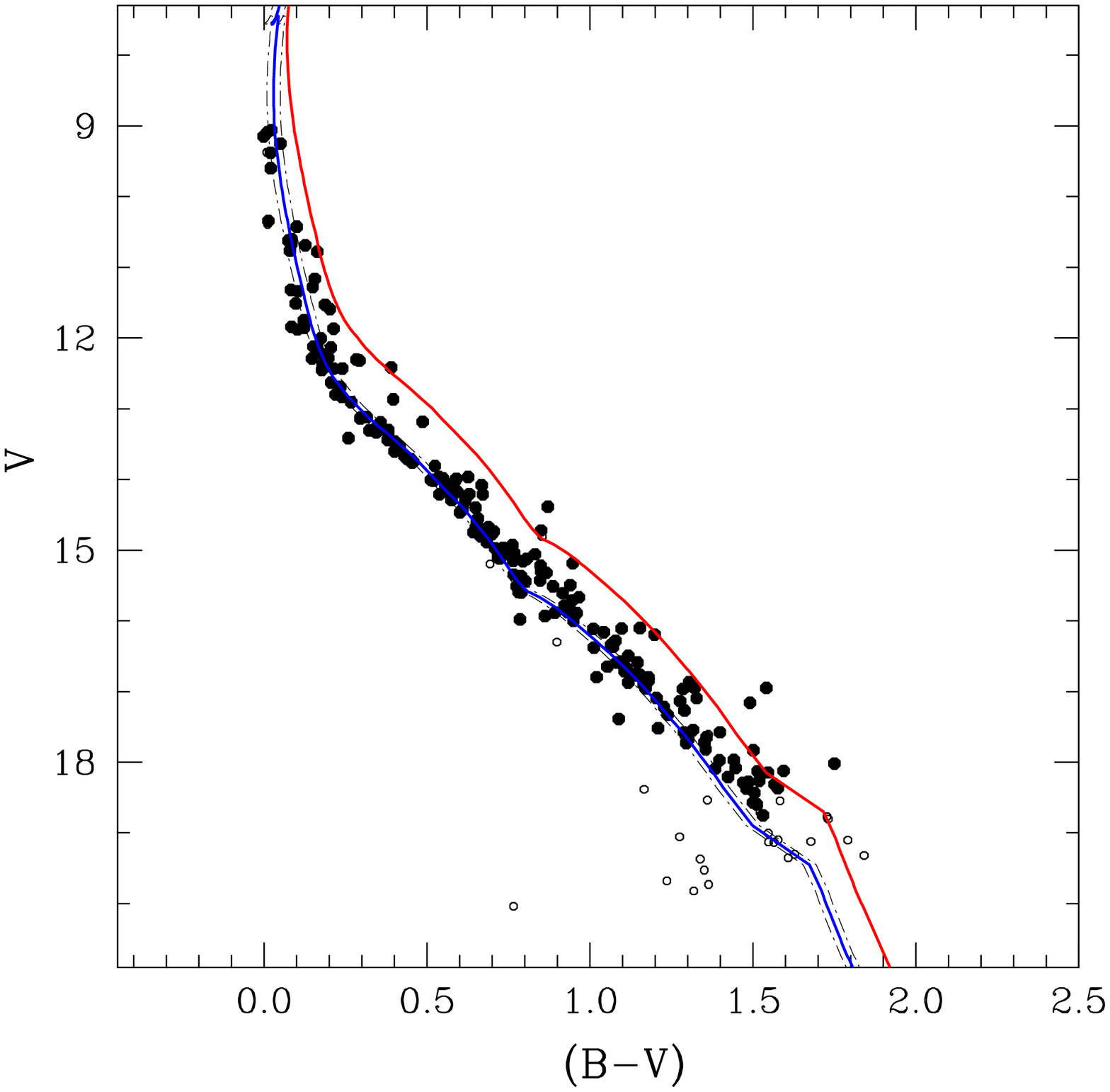}}
\caption{The lower and upper panels show $(B-V)/V$ and $(V-I)/V$ CMDs for cluster members, respectively. The open circles represent stars containing photometric errors $eV>0.02$ mag. The solid lines represent blue and red sequences of the best fit solar metallicity isochrones for $\log$(Age)=7.44$\pm$0.02. Here, distance modulus is taken as $(m-M)_0=10.33\pm0.11$ obtained through the Gaia DR2 parallaxes and reddening $E(B-V)=0.24\pm0.02$ from the $(U-B)/(B-V)$ diagram. The two dashed black lines around the blue sequence represent the error of 0.02 mag in the reddening $E(B-V)$.}
\label{col_mag}
\end{figure}

Colour-magnitude diagrams (CMDs) are the most effective tool to determine distance and age of the cluster provided we know the reddening in the direction of the cluster. As we have already determined precise distance of the cluster from the parallax measurements of the cluster members and reddening $E(B-V)$ and $E(V-I)$ from the TCDs, we now turn to the determination of age of the cluster through comparison of observed CMDs to the theoretical isochrones. In Figure~\ref{col_mag}, we draw simultaneous $(B-V)/V$ and $(V-I)/V$ CMDs. Here, we do not consider those stars that have large photometric errors of $eB>0.10$ mag and $eV>0.05$ mag. The main sequence of the cluster is clearly evident in both CMDs which also implies that the membership selection  based on the kinematic measurements is quite robust. We overplot Marigo's theoretical isochrones for solar metallicity \citep{2017ApJ...835...77M} on the CMDs by varying age simultaneously in both CMDs while keeping reddening $E(B-V)=0.24$ mag and $E(V-I)=0.30$ mag fixed as determined in Section~\ref{red_opt}. The distance modulus is also kept fixed to $(m-M)_0=10.33$ mag as obtained through Gaia DR2 parallaxes of the member stars. From the best visual isochrone fit to our CMDs with varying age to the blue edge of the stellar population of the main-sequence stars, we obtained $log(Age/yr)=7.44\pm0.02$ for the cluster NGC\,1960. The CMDs show a well populated but broad main sequence that may be due to the presence of binary stars within the cluster or variable reddening which we have also noticed through the scattering in $(B-V)/(U-B)$ diagram in Section~\ref{red_opt}. Therefore, to illustrate the binary effect, we also draw the same isochrones through red sequence by shifting 0.75 mag in $V$ magnitude and 0.042 in $(B-V)$ and $(V-I)$ colours in comparison of their blue sequences.

%---------------------------- Fig. BV/V --------------------
%------------------------------------ End Fig. ----------------------

Based upon photometric observations, the age of the cluster NGC\,1960 has been earlier estimated as 25.1 Myr by SHA06, \citet{2009MNRAS.399.2146W} and $26.3^{+3.2}_{-5.2}$ Myr by \citet{2013MNRAS.434..806B} which are consistent with the present estimate of $27.5^{+1.3}_{-1.2}$ Myr. However, \citet{2000A&A...357..471S} determined a relatively smaller age ($16.3^{+10}_{-5}$ Myr) for this cluster albeit with very large uncertainty. Though isochrone fitting is often used to estimate the age of the cluster in the absence of more valuable but lesser available spectroscopic observations but it should be kept in mind that determining precise age through isochrone fitting in clusters, when no evolved stars are found, is relatively difficult as it lends a larger uncertainty \citep[e.g.,][]{2001MNRAS.327...23S}. Using the spectroscopic data, \citet{2013MNRAS.434.2438J} determined age of this cluster as log(t/yr)=7.34$\pm$0.08 ($\sim 22\pm4$) for the cluster through the luminosity of the stars that have not yet consumed their lithium. Their age estimation is though slightly smaller in comparison of the other estimates which is not surprising considering lithium depletion boundary (LDB) is quite sensitive to the choice of evolutionary models \citep{2013MNRAS.434.2438J}. For example, \citet{2015ApJ...813..108D} reported an age of 112$\pm$5 for the Pleiades cluster using the LDB method which is well below the commonly found age of $\sim$ 125 Myr for this young cluster. Similarly, in a recent study by \citet{2018ApJ...856...40M} on the Haydes cluster applying the same approach, they provided a range of ages between 440 to 940 Myrs for the Haydes employing four different evolutionary models.

%
%--------------------------------- Fig. TCD ----------------------
\begin{figure}
\centerline{\includegraphics[width=8.5cm,height=11.5cm]{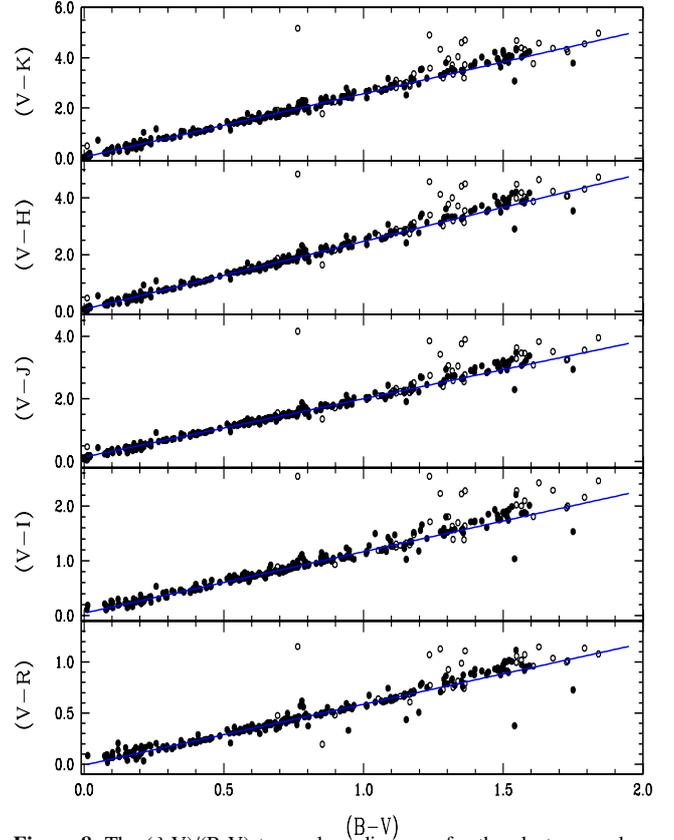}}
\vspace{-0.6cm}
\caption{The ($\lambda$-V)/(B-V) two-colour diagrams for the cluster members where $\lambda$ is $R$, $I$, $J$, $H$, and $K$ bands from the bottom to top panels. The open circles represent stars having photometric errors $eV>0.02$ mag. The thick continuous lines represent best fit slopes in each plot.}
\label{tcd_5col}
\end{figure}
%--------------------------------- End Fig. -------------------------
%
\subsection{Extinction law}\label{elaw}
Generally, normal reddening law is applicable when dust and intermediate stellar gases are absent in the line of sight of the cluster \citep{1978APJ...223..168S}. However, reddening law is expected to be different in the presence of dust and gas. The $(V-\lambda)/(B-V)$ TCDs have been widely used to see the influence of the extinction generated by the diffuse interstellar material from that of the intra-cluster medium \citep{1990A&A...227..213C}. We investigated the nature of reddening law towards the cluster direction using $(V-\lambda)/(B-V)$ TCDs, where $\lambda$ is $R$, $I$, $J$, $H$, and $K$ bands. We illustrate the $(V-\lambda)/(B-V)$ diagrams for the cluster members in Figure~\ref{tcd_5col}. A best linear fit in the TCD for the cluster members gives the slope ($m_{cluster}$) in the corresponding TCD. The resultant values of the $m_{cluster}$ are listed in Table~\ref{tcd} along with their normal values. Our slopes are quite comparable with those obtained for the diffuse interstellar material which suggests a normal reddening law in the direction of the cluster region.

%TCD slopes
  \begin{table}
   \small
    \caption{The slopes of $(V-\lambda)/(B-V)$ two-colour diagrams and total-to-selective extinction $R_{cluster}$ in the direction of the cluster.}
    \label{tcd}
    \begin{center}
      \begin{tabular}{cccc} \hline
      TCD               &    $m_{cluster}$ &  $m_{normal}$ & $R_{cluster}$ \\  \hline \vspace{0.15cm}
      $\frac{V-R}{B-V}$ &   0.59$\pm$0.01  &      0.55     &  3.33 \\          \vspace{0.15cm}
      $\frac{V-I}{B-V}$ &   1.12$\pm$0.01  &      1.10     &  3.16 \\          \vspace{0.15cm}
      $\frac{V-J}{B-V}$ &   1.86$\pm$0.01  &      1.96     &  2.94 \\          \vspace{0.15cm}
      $\frac{V-H}{B-V}$ &   2.39$\pm$0.01  &      2.42     &  3.06 \\          \vspace{0.15cm}
      $\frac{V-K}{B-V}$ &   2.52$\pm$0.01  &      2.60     &  3.00 \\  \hline
      \end{tabular}
    \end{center}
  \end{table}

A total-to-selective extinction $R_{cluster}$ is determined using the relation given by \citet{1981A&AS...45..451N} as
$$
R_{cluster}=\frac{m_{cluster}}{m_{normal}} \times R_{normal}.
$$
$R_{normal}$ is known to be correlated with the average size of the dust grains causing the extinction. The typical value for $R_{normal}$ is 3.1 for the diffuse interstellar material in our Galaxy \citep{1989ApJ...345..245C}. Using $R_{normal} = 3.1$, we determined $R_{cluster}$ in these five colours and given in Table~\ref{tcd}. The global mean value of $R_{cluster}$ is estimated to be 3.10$\pm$0.06 for the cluster NGC\,1960 which is in excellent agreement with the normal extinction law in the direction of the cluster. This suggests that there is no discernible anomalous reddening in the cluster region.

The basic parameters of the cluster NGC\,1960 derived in the present study are summarized in Table~\ref{parameters}.   
%
%Physical parameters of cluster
 \begin{table}
    \caption{The basic physical parameters for NGC~1960 as determind in the present study.}
    \label{parameters}
    \centering
      \begin{tabular}{cc} \hline
Cluster parameters     &       Values             \\ \hline
$N_{cluster}$                 & 262               \\
$\bar{\mu}_{\alpha}$ (mas/yr) &  -0.143$\pm$0.008 \\
$\bar{\mu}_{\delta}$ (mas/yr) &  -3.395$\pm$0.008 \\
$\overline\omega$ (mas)       &   0.831$\pm$0.048 \\
$E(B-V)$(mag)            &  0.24$\pm$0.02    \\ 
Total-to-selective extinction &  $3.10\pm0.06$    \\
$(m-M)_0$ (mag)               & 10.33$\pm$0.11    \\
Distance (kpc)                &  1.17$\pm$0.06    \\ 
Diameter (pc)                 &  9.80$\pm$0.02    \\ 
log(Age/yr)                   &  7.44$\pm$0.02    \\
\hline 
  \end{tabular}
  \end{table}

\section{Dynamical study of the cluster}\label{dynamical}
In order to understand the dynamical behaviour of the cluster NGC\,1960, the luminosity function, mass function and mass segregation process are examined in the following sub-sections.
\subsection{Luminosity function} \label{vlf}
The luminosity function (LF) is defined as the total number of cluster members in different magnitude bins. However, estimation of LF using a complete sample of cluster stars is not straight forward, as many biases and uncertainties are involved in its determination \citep[e.g.,][]{2012ARA&A..50...65L, 2014prpl.conf...53O}. For example, the data incompleteness increases with the fainter magnitudes. However, it is better than 90\% up to 19 mag in SHA06 and it is further improved in the present catalogue as we added more photometric data to the central region of the cluster. Hence we assumed that the photometry presented here are not affected by the data incompleteness for the stars brighter than 19 mag which we considered to determine the LF. We found 229 cluster members between the magnitude limit 9 to 19 mag in the $V$-band. The LF was estimated in a bin width of 1 mag. We estimated the mass of each star photometrically by comparing its colour and magnitude from the theoretical isochrones of solar metallicity \citep{2017ApJ...835...77M} for the estimated log(t/yr) = 7.44, extinction $E(B-V)$ = 0.24 mag, and $(m-M)_0 = 10.33$ mag. The mass for each cluster member was determined from its nearest neighbour on the selected isochrones. In Table~\ref{LF}, we provide mass range, mean mass and cluster members in different brightness range for the cluster.
%
%Luminosity and Mass functions
%----------------------------------------------------------------------------
\begin{table}
\centering
\caption{Luminosity and Mass functions in the cluster NGC~1960 for the stars $V\le19$ mag. Here, $\phi = dN/d\,log(\bar{m})$ is the number of stars per unit logarithmic mass.}
\label{LF}
\begin{tabular}{ccccccc }
\hline
$V$ range& Mass range  & $N_c$ & $\bar{m}$    & $log(\bar{m})$ & $log(\phi)$ & $\sigma[log(\phi)]$ \\
(mag)    &($M_{\odot}$)&       & ($M_{\odot}$)&                &             &          \\
\hline                         
 9-10    & 7.32-5.66   &    8  &  6.56        &   0.817        &     1.854   &      0.354    \\
10-11    & 5.64-4.00   &    9  &  4.83        &   0.684        &     1.781   &      0.333    \\
11-12    & 3.98-2.76   &   12  &  3.29        &   0.518        &     1.880   &      0.289    \\
12-13    & 2.75-1.93   &   21  &  2.23        &   0.348        &     2.137   &      0.218    \\
13-14    & 1.92-1.52   &   28  &  1.72        &   0.236        &     2.437   &      0.189    \\
14-15    & 1.52-1.26   &   40  &  1.40        &   0.147        &     2.708   &      0.158    \\
15-16    & 1.26-0.99   &   38  &  1.10        &   0.041        &     2.565   &      0.162    \\
16-17    & 0.99-0.89   &   33  &  0.94        &  -0.027        &     2.866   &      0.174    \\
17-18    & 0.89-0.79   &   23  &  0.85        &  -0.072        &     2.668   &      0.209    \\
18-19    & 0.79-0.72   &   25  &  0.76        &  -0.120        &     2.805   &      0.200    \\
\hline
\end{tabular}
\end{table}

\subsection{The present-day mass function} \label{lmf}
\begin{figure}
%\vspace{-1.4cm}
\centerline{\includegraphics[width=8.5cm,height=6.5cm]{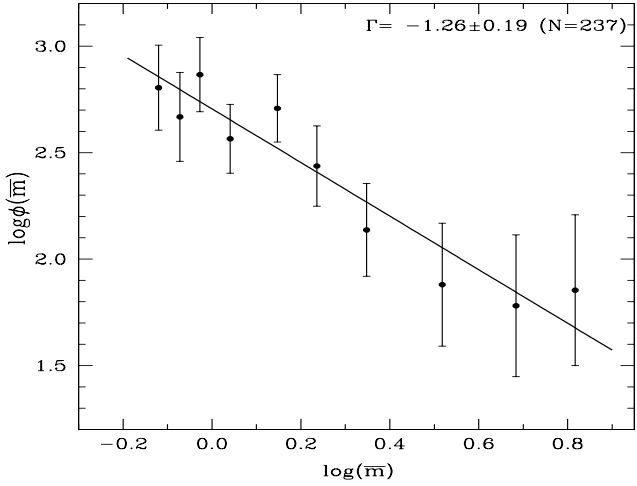}}
%\vspace{-0.5cm}
\caption{MF derived for the cluster NGC\,1960. The error bars represent $1/\sqrt N$ errors. The continuous line is the best fit to the mass range $0.72 \le \frac{M}{M_{\odot}} \le 7.32$. The estimated MF slope is given at the top of the plot.}
\label{mf}
\end{figure}

The initial mass function (IMF), i.e. the frequency distribution of stellar masses at the time of birth, is a fundamental parameter in the study of star formation and evolution in the cluster. It represents the distribution of stellar masses per unit volume in a star formation event and knowledge of IMF is very effective to determine the subsequent evolution of the cluster. The direct measurement of IMF is not possible due to dynamical evolution of stellar systems though we can estimate the present-day mass function (MF) of the cluster. Since the age of the cluster NGC\,1960 is relatively young ($\sim$ 27 Myr), the present-day MF can be considered as IMF \citep{2008MNRAS.386.1380K}. The MF is defined as the relative numbers of stars per unit mass and can be shown by a power law $N(\log M) \propto M^{\Gamma}$. The slope, $\Gamma$, of the MF can therefore be determined as
$$
\Gamma = \frac{d\log N(\log\it{m})}{d\log\it{m}}
$$
where $N\log(m)$ is the number of stars per unit logarithmic mass. In last two columns of Table~\ref{LF}, we provide MF and corresponding error for different magnitude bins. The MF in the cluster fitted for the main-sequence stars with masses $0.72 \le M/M_{\odot} \le 7.32$ is shown in Figure~\ref{mf}. The error bars are determined assuming Poisson statistics which shows considerably large value due to lower number of stellar counts in each bin. The MF slope is found to be -1.26$\pm$0.19 with a Pearson correlation coefficient of 0.96. The quoted uncertainty in the MF slope come from the linear regression solution in the fit. In the present estimation of MF, the effect of field star contamination is considered to be negligible due to fact that only cluster members are used in the study. Our estimated MF slope is in excellent agreement with the Salpeter MF slope of -1.35 \citep{1955ApJ...121..161S}. However, \citet{2008AJ....135.1934S} reported a stepper MF slope ($\Gamma$=$-1.80{\pm}0.14$) in this cluster for the narrower mass range of $1.01 < M/M_{\odot} < 6.82$ relying on the approach based on the statistical subtraction of stars to find the probable cluster members.

%
%--------------------------Fig. Mass function----------------------
%-----------------------------------------------------------
%

\begin{figure}
\centering
\centerline{\includegraphics[width=8.5cm,height=11.0cm]{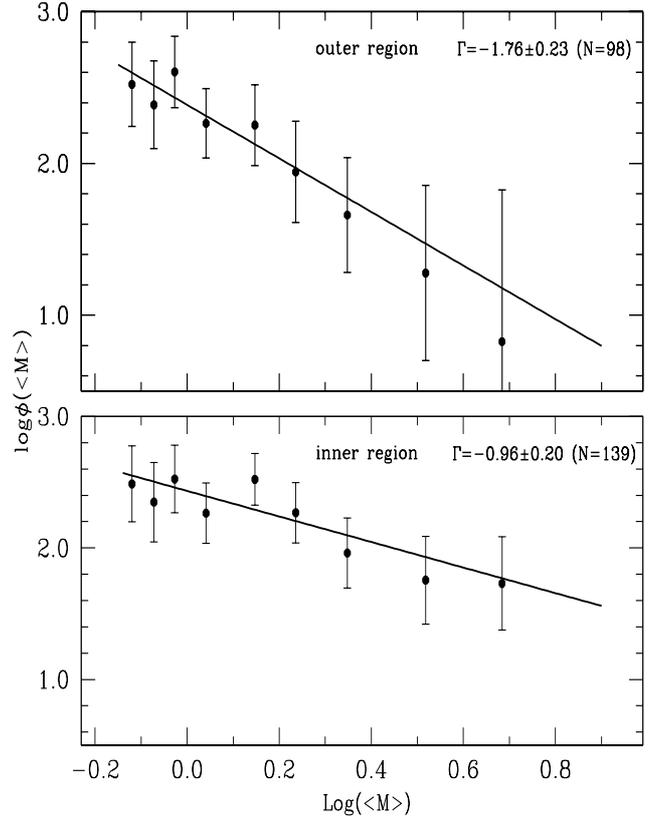}}
%\vspace{-0.5cm}
\caption{Same as Figure~\ref{mf} but for the regions between 0-5 arcmin (inner region) and 5-14 arcmin (outer region) from the cluster center.}
\label{radial_MFslope}
\end{figure}

To further probe the mass segregation in the cluster, we investigated the radial variation in the mass function slope. We determined MF values for two separate regions containing central region up to $5^\prime$ from the cluster center and outer region in between $5^\prime$ to $14^\prime$ which contains 138 and 97 stars, respectively. In Figure~\ref{radial_MFslope}, we illustrate MF variation in these two regions separately. The estimated slopes of the MFs in the inner and outer regions are given at the right corner of each plot in Figure~\ref{radial_MFslope}. The mass function slopes differ from each other by more than 1$\sigma$, where $\sigma$ is the error associated with the slope. As can be seen from the figure, the MF slope in the inner region is clearly flatter than the outer region while overall MF slope is in remarkable agreement with the Salpeter value. This again suggests that the mass segregation process is taking place in the cluster.
%
%-------------------------------Fig. MF slope -----------------
%-------------------------------------------------------------
%
\subsection{Dynamical evolution}\label{dyna_evol}
The dynamical evolution of a star cluster is primarily characterised by the mass segregation, tidal radius, crossing time, and relaxation time which are briefly discussed below.
\subsubsection{Mass segregation}\label{segre}
The dynamical evolution gradually drives the system towards equipartition resulting the low mass stars attaining higher velocities hence occupying larger orbits around the cluster center \citep{1986AJ.....92.1364M}. This process, commonly known as mass segregation, results in accumulating more massive stars to the core and low-mass stars to the peripheral region of the cluster. Finally, low-mass cluster members, which acquire large enough velocity from the equipartition of energy, are escaped away from the cluster's tidal field resulting change in the morphology of the spatial mass distribution in the cluster \citep{1988MNRAS.234..831S}. 

\begin{figure}
\centering
\centerline{\includegraphics[width=8.5cm,height=6.5cm]{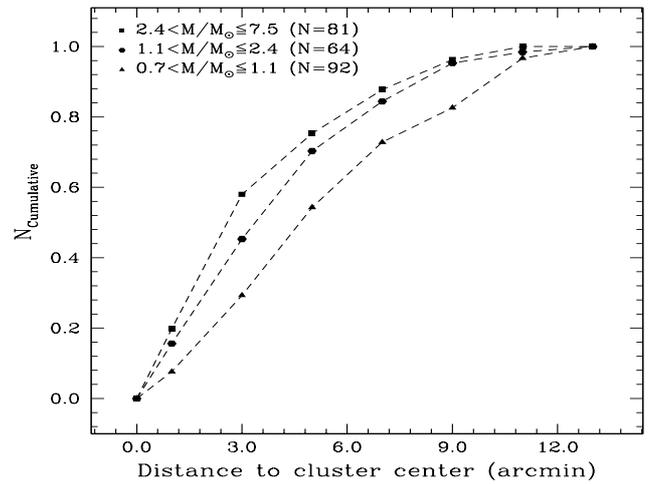}}
\caption{The variation of cumulative distribution of cluster members in radial bins relative to the cluster center. The distribution is estimated for three different mass ranges as given at the top of the plot along with total number of cluster members in each mass bin.}
\label{mass_segg}
\end{figure}

To study the mass segregation in the cluster NGC\,1960, we draw the variation of cumulative number in Figure~\ref{mass_segg} for the cluster members along the radial distance in three different mass bins of $M/M_\odot\le1.1$ (92 stars), $1.1<M/M_\odot\le1.5$ (64 stars), and $M/M_\odot>1.5$ (81 stars). The mass ranges are selected in order to get enough statistical sample in each bin. From the radial variations of stars in these three mass bins, it is quite evident that the massive stars are dominant in the core of the cluster while low mass stars are distributed in the outer region. This result agrees with the theoretical expectations of mass segregation effect within the cluster. We also performed Kolmogorov-Smirnov (K-S) test of these distributions to examine whether they are statistically different or not and conclude with 95$\%$ confidence level that mass segregation effect in the cluster NGC\,1960 is present.
%
%----------------------------- Fig. Mass Segregation---------
%----------------------------------------------------------
% 
%
\subsubsection{Tidal radius}\label{Rt}
The study of tidal interactions in the open clusters plays an important role in understanding the initial structure and dynamic evolution of the clusters \citep{2010MNRAS.402.1841C, 2015MNRAS.449.1811D}. The tidal radius is described as the radial distance from the cluster center where gravitational acceleration caused by the cluster is almost equal to the tidal acceleration caused due to the Galaxy \citep{1957ApJ...125..451V}. It is believed that the stars are generally gravitationally bound to the cluster within the tidal radius due to effective potential of the cluster. \citet{2000ApJ...545..301K} provided following relation to determine tidal radius
$$
R_{t} = \left(\frac{M_{C}}{2 M_{G}}\right)^{1/3}\times R_{G}
$$
Where $R_t$ is the tidal radius, $M_C$ is the total cluster mass and $M_{G}$ is the mass of the Galaxy within the Galactocentric radius $R_G$. Given the uncertainties in the actual masses and extent of the clusters, tidal radii are often poorly determined. Using the cluster Galactic positions and distance, we obtained the Galactocentric distance of the cluster NGC\,1960 as $R_G = 9.25$ kpc. Here, $R_\odot$, the distance between the Sun and the Galactic Centre, is considered as 8.0$\pm$0.3 kpc \citep{2012PASJ...64..136H, 2018PASP..130b4101C}. It is difficult to estimate an accurate value of $M_C$ without identifying all the low-mass members. Nonetheless, we can still make a relatively good estimate from our data because we identify cluster members down to $V=20.65$ mag having mass as low as 0.49 $M_\odot$. Using the 262 main-sequence stars within $0.49 \leq M/M_\odot \leq 7.32$ obtained in the present analysis, $M_C$ is estimated to be $\sim 416.7~M_\odot$ which yields a mean stellar mass of $\sim 1.6~M_\odot$. Employing the relation given by \citet{1987ARA&A..25..377G}, $M_{G}$ was estimated to be $\sim$ 1.57$\times$10$^{11}~M_\odot$. Using the above given relation, we obtained the tidal radius as 10.1 pc. Any star beyond this radius would be gravitationally unbound to the cluster NGC\,1960. \citet{2008A&A...477..165P} reported a slightly larger value of tidal radius as $R_{t} = 10.6\pm1.6$ pc for the cluster. This is well understood since they used a larger Galactocentric distance for the Sun which consequently has increased their assessment of tidal radius.
\subsubsection{Cluster half-radius and Crossing time}\label{Rh}
The cluster half-radius, R$_{h}$, is defined as the radius within which half of the total cluster mass lies. Sometimes it is also called half-mass radius. To determine R$_{h}$, we estimated cumulative mass of the stars by increasing radial distance from the cluster center. We select the radial distance where we found half of the total cluster mass. This results a R$_{h}$ of 6.5 arcmin for the cluster which corresponds to a linear radius of 2.26 pc. As we earlier obtained a cluster radius of 4.90 pc for NGC\,1960, the cluster half-radius is found to be slightly smaller but comparable in comparison of the half of the cluster linear radius. 

The crossing time ($t_{cr}$) defined as a time in which a star with a typical velocity travels through the cluster under the assumption of virial equilibrium:
$$
t_{cr} = \sqrt{\frac{2R_h^3}{GM_C}}
$$
\citep[e.g.,][]{1987degc.book.....S, 2005A&A...429..173L} where $R_h$ is the half-mass radius, $M_c$ is the total cluster mass and $G$ is the gravitational constant. We determined $t_{cr}$=3.50 Myr for the cluster NGC\,1960. For a cluster having an age of $\sim$ 27.5 Myr, this corresponds to $\approx$ 8 crossings since the formation of the cluster. It is generally believed that after several crossing times, the cluster obtains a virial equilibrium \citep[e.g.,][]{2012MNRAS.421.3338A} and becomes dynamically relaxed.
\subsubsection{Relaxation time}\label{drt}
The dynamical relaxation time, $T_E$, is the time in which individual cluster members exchange energies and their velocity distribution approaches Maxwellian equilibrium. The $T_E$ corresponds to the time over which the cumulative effect of stellar encounters becomes comparable to the star's velocity itself \citep{2019arXiv191104562D} and can be expressed as
$$
T_E = \frac{0.89 \times (N R_h^3/\bar{m})^{1/2}} {\ln(0.4N)}
$$
where $N$ is the total number of cluster members, $R_h$ is the half-radius (in parsecs), $\bar{m}$ is the mean stellar mass (in solar units) and $T_E$ is the relaxation time in Myr \citep[cf.,][]{1971ApJ...164..399S}. Considering the mean stellar mass of 1.6 M$_\odot$ and cluster half radius of 2.26 pc, we obtained dynamical relaxation time $T_E$ = 19.2\,Myr for the cluster NGC\,1960 which is very close to the  present cluster age of about 27\,Myr. The star clusters generally become dynamically relaxed after few relaxation times \citet{2008AJ....135.1934S}. However, a  half-mass radius comparable to the half of the cluster radius besides relaxation time comparable to present cluster age suggest that NGC\,1960 is not completely dynamically relaxed as yet and mass segregation is still an ongoing process in the cluster.

A summary of the parameters obtained in the present study from the dynamical study of the cluster NGC\,1960 is given in Table~\ref{dyna}.
%
%Dynamical parameters
% tab:dyna
\begin{table}
\centering
\caption{Parameters determined from the dynamical study of cluster NGC~1960.}
\label{dyna}
\begin{tabular}{ll}
\hline
Total cluster mass ($M_C$)       & 416.7 $M_\odot$   \\
Mean stellar mass ($\bar{m}$)    & 1.6 $M_\odot$  \\
Cluster half-mass radius ($R_h$) & 2.26 pc         \\
Tidal radius ($R_t$)             & 10.1 pc         \\
Relaxation time ($T_E$)          & 19.2 Myrs       \\
Crossing time ($T_{cr}$)         & 3.3 Myrs        \\
\hline
\end{tabular}
\end{table}

\section{Identification of Variable stars} \label{vs}
As stated in Section~\ref{data}, we accumulated 235 frames in the $V$ band on 43 nights for the central $13^\prime \times 13^\prime$ field of the cluster NGC\,1960. We used this time-series photometric data to search for the variable stars in the cluster. Though we identified 3962 stars within $14^\prime$ radius of the cluster, we found only 1386 stars in the central $13^\prime \times 13^\prime$ target field of the ST. It should be noted here that due to varying sky conditions during various observing runs, and different exposure times in different frames, not all the stars could be identified in all the frames. As we have carried out absolute photometry on the night of 30 November 2010, we converted instrumental magnitudes of the stars into the corresponding absolute magnitudes on each night by applying the necessary photometric corrections as following.
$$ V = a \times v + b$$
where $V$ and $v$ are the photometric magnitude on the night of standardization and instrument magnitude of the same star on the target frame. In each frame, we considered more than 100 stars. The coefficients $a$ and $b$ in each frame were calculated by a least square linear fit using the non-saturated stars brighter than 15 mag. Here, the color term was not used as it was found to be insignificant. We then searched for the stellar variability in 1386 stars by looking for the magnitude variations over entire monitoring period.

It was noticed that due to changes in observing conditions during our observations, there was a large variation in the data quality of the photometric light curves. Before analysing variable stars, we therefore carefully eliminated possible outliers from the photometric data. In few cases, some points were removed on the basis of extreme excursions from the mean value. We considered only those stars which fall within 10 pixels from the edge of the target image and present in more than 50 images.
\subsection{Periodic variables} \label{pv}
In order to search for periodic variables, the time-series $V$ band magnitudes of all the 1386 stars were subjected to the periodicity analysis. We used Lomb-Scargle algorithm \citep{1976Ap&SS..39..447L, 1982ApJ...263..835S} within the software PERIOD04 \citep{2005CoAst.146...53L}, especially dedicated to the statistical analysis of large astronomical time series data containing gaps and to extract the individual frequencies from the multi-periodic content of the time-magnitude variation. This method computes the Fourier power spectrum by fitting sine and cosine terms over a large number of frequencies in the given frequency range. Only stars that lied within $V<19.5$ were considered for the variability search since photometric magnitudes have relatively large errors towards the fainter end. It is noticed that many spurious variables were also detected with periods in harmonics of 1 sidereal day i.e. at period of 1 sidereal day/n where n=2,3,4,.... We also did not consider those periodic variation where amplitude variation was smaller than the mean uncertainty in the data points. 

Phase for each potential variable star was calculated using the following equation:
$$Phase = \left(\frac{JD - JD_0}{P}\right)  - INT\left(\frac{JD - JD_0}{P}\right)$$
Where $JD$ is the time of observation, $JD_0$ is an arbitrary epoch of observation. All the periodic variables were checked by reviewing the phase folded light curves created with their periods. The light curves of only those variables were identified that showed a good periodic brightness variation over the entire phase.

%
%----------------------Fig: Periodic Variables  ----------------------
\begin{figure*}
\vspace{-0.5cm}
\centering
\centerline{\includegraphics[width=18.0cm,height=24.0cm]{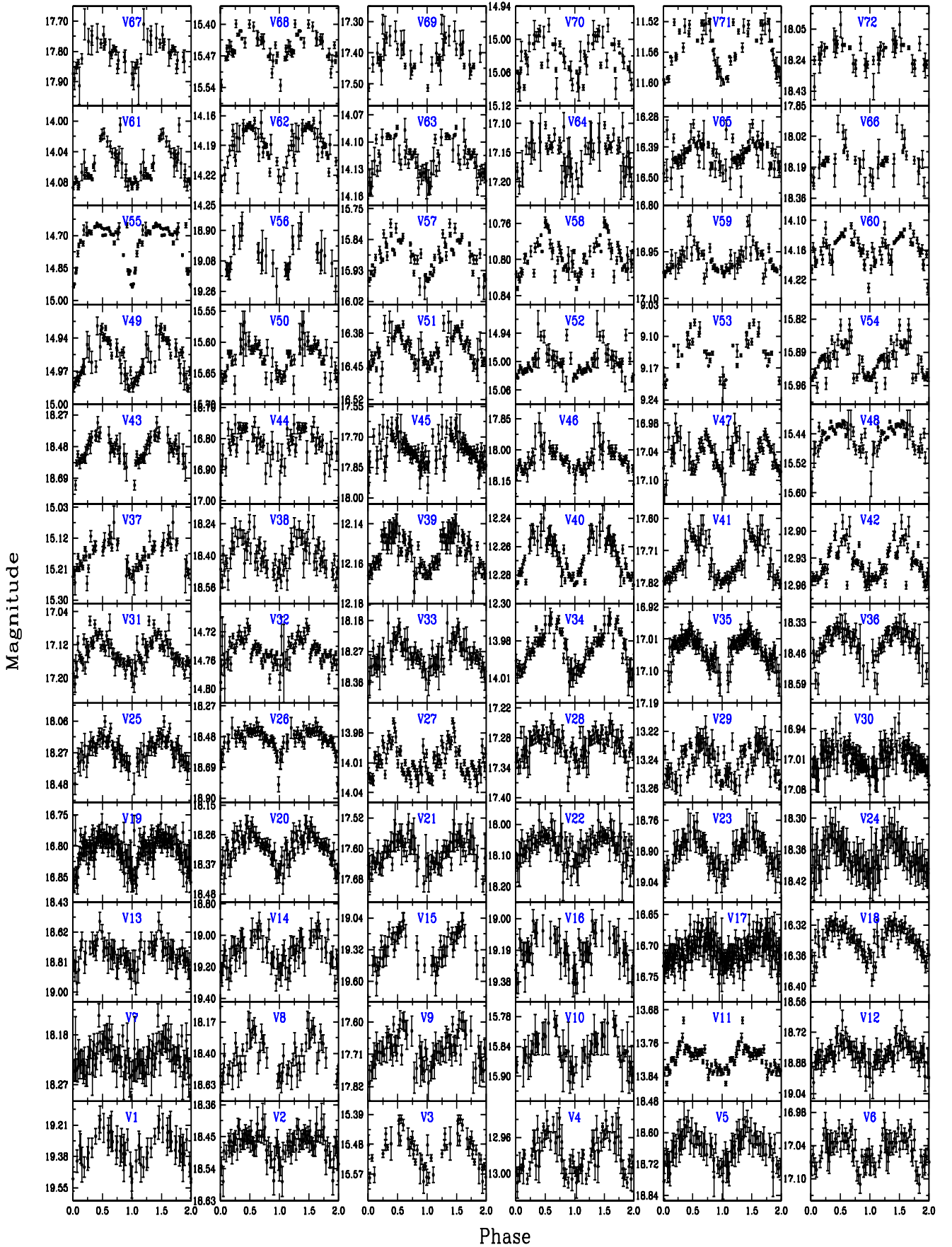}}
%\vspace{-0.5cm}
\caption{The phased light curves of 72 periodic variables found in the field of NGC\,1960. The star ID is given at the top of each individual light curve. Phase is plotted twice and in such a way that the minimum brightness falls near to zero phase.}
\label{periodic}
\end{figure*}
%----------------------------- End Fig ----------------------
%
% List of periodic variables
\begin{table*}
\caption{The details of the 72 periodic variables found in the cluster NGC~1960. The variable stars identification are sorted in the order of increasing period. The columns give variable sequence, cluster ID, X, Y, RA, DEC, $V$ band mean magnitude, ($B-V$) and ($V-I$) colours, period, uncertainty in period, amplitude of variation in $V$ band, and epoch of minimum light (JD-2450000). Last column gives the membership status of the variable star.}
\centering
\label{table_periodic}
\scriptsize
\begin{tabular}{ccccccccccccc}
\hline
No. & ID   &  RA (J2000) & DEC (J2000)  &   V    & (B-V)  & (V-I)   &   Period &$\sigma$(Period)& A$_V$ & N&   $T_0$     & Membership \\
    &      & (hh:mm:ss)  & (dd:mm:ss)   & (mag)  & (mag)  & (mag)   &   (day)  &  (day)   & (mag) &     &   (day)       &     \\
\hline
V01 & 2549 & 05:35:50.59 &  +34:12:05.7 & 19.338 &  1.875 &   2.258 &  0.02856 &  0.00001 &  0.17 &  45 &   5531.34792   &  Field    \\
V02 & 1671 & 05:36:28.94 &  +34:04:44.9 & 18.472 &  1.139 &   1.356 &  0.04592 &  0.00067 &  0.09 & 147 &   5129.38877   &  Field    \\
V03 &  420 & 05:35:42.66 &  +34:05:28.4 & 15.501 &  0.751 &   0.897 &  0.05045 &  0.00001 &  0.09 &  46 &   5134.32933   &  Field    \\
V04 &  114 & 05:36:03.57 &  +34:15:36.3 & 12.979 &  0.263 &   0.345 &  0.05763 &  0.00001 &  0.04 &  73 &   5134.33068   &  Field    \\
V05 & 1918 & 05:35:49.30 &  +34:14:48.3 & 18.661 &  1.273 &   1.487 &  0.06214 &  0.00001 &  0.12 &  69 &   5247.07428   &  Field    \\
V06 &  890 & 05:36:22.17 &  +34:01:53.9 & 17.047 &  1.022 &   1.283 &  0.06487 &  0.00001 &  0.06 &  50 &   5201.11301   &  Field    \\
V07 & 1469 & 05:36:19.69 &  +34:04:49.5 & 18.219 &  1.067 &   1.211 &  0.06930 &  0.05309 &  0.09 & 161 &   5129.38162   &  Field    \\
V08 & 1287 & 05:36:08.85 &  +34:11:40.0 & 18.423 &   -    &   1.270 &  0.07584 &  0.00001 &  0.23 &  41 &   5143.29719   &  Field    \\
V09 & 1184 & 05:36:47.28 &  +34:09:28.5 & 17.691 &  1.128 &   1.288 &  0.10021 &  0.00001 &  0.11 &  60 &   5129.35413   &  Field    \\
V10 &  535 & 05:36:30.78 &  +34:15:20.3 & 15.849 &  1.845 &   2.056 &  0.10036 &  0.00001 &  0.06 &  45 &   5134.27946   &  Field    \\
V11 &  181 & 05:36:36.73 &  +34:14:29.1 & 13.803 &  0.695 &   0.972 &  0.10375 &  0.00001 &  0.08 & 127 &   5129.38467   &  Field    \\
V12 & 2052 & 05:35:52.85 &  +34:08:01.2 & 18.832 &  1.143 &   1.346 &  0.11639 &  0.00046 &  0.16 & 118 &   5531.31584   &  Field    \\
V13 & 1978 & 05:35:46.69 &  +34:02:09.6 & 18.756 &  1.214 &   1.424 &  0.11703 &  0.00001 &  0.19 &  52 &   5868.37142   &  Field    \\
V14 & 2000 & 05:36:16.59 &  +34:07:10.6 & 19.095 &   -    &   1.833 &  0.12555 &  0.00002 &  0.20 &  60 &   5853.39800   &  Field    \\
V15 & 2542 & 05:35:57.89 &  +34:14:16.4 & 19.303 &  1.383 &   1.811 &  0.12789 &  0.00002 &  0.28 &  46 &   5531.34199   &  Field    \\
V16 & 2338 & 05:35:53.24 &  +34:15:34.0 & 19.200 &  1.365 &   1.426 &  0.13360 &  0.00002 &  0.19 &  42 &   5895.17996   &  Field    \\
V17 &  779 & 05:35:47.92 &  +34:05:13.8 & 16.710 &  0.989 &   1.074 &  0.14993 &  0.00002 &  0.05 & 188 &   5129.25166   &  Field    \\
V18 &  670 & 05:35:48.00 &  +34:12:41.1 & 16.341 &  0.699 &   0.841 &  0.15751 &  0.00013 &  0.04 & 174 &   5134.30299   &  Field    \\
V19 &  828 & 05:35:58.87 &  +34:08:25.4 & 16.804 &  1.170 &   1.456 &  0.17074 &  0.00003 &  0.05 & 188 &   5129.35734   &  Field    \\
V20 & 1632 & 05:36:18.71 &  +34:06:40.1 & 18.299 &  1.213 &   1.393 &  0.17256 &  0.00003 &  0.11 & 164 &   5129.36725   &  Field    \\
V21 & 1189 & 05:36:46.09 &  +34:08:58.5 & 17.605 &  1.050 &   1.339 &  0.17794 &  0.00003 &  0.08 &  93 &   5129.40036   &  Field    \\
V22 & 1447 & 05:35:44.53 &  +34:05:33.3 & 18.070 &  1.116 &   1.277 &  0.18175 &  0.00003 &  0.10 & 121 &   5134.24420   &  Field    \\
V23 & 2102 & 05:36:17.38 &  +34:07:10.9 & 18.907 &  1.524 &   1.699 &  0.18471 &  0.00082 &  0.14 & 115 &   5531.19240   &  Field    \\
V24 & 1702 & 05:36:15.96 &  +34:09:58.4 & 18.371 &  1.506 &   1.563 &  0.18769 &  0.00054 &  0.06 & 151 &   5144.33698   &  Field    \\
V25 & 1387 & 05:36:01.96 &  +34:09:53.7 & 18.242 &  1.388 &   1.539 &  0.19029 &  0.00112 &  0.21 &  84 &   5129.40797   &  Field    \\
V26 & 1751 & 05:35:56.16 &  +34:14:01.7 & 18.494 &  1.022 &   1.657 &  0.19421 &  0.00162 &  0.21 & 136 &   5129.24468   &  Field    \\
V27 &  192 & 05:36:41.30 &  +34:09:29.4 & 14.007 &  0.590 &   0.619 &  0.20821 &  0.00016 &  0.03 & 212 &   5129.08380   &  {\bf Cluster}  \\
V28 & 1031 & 05:35:54.43 &  +34:11:27.8 & 17.295 &  1.056 &   1.617 &  0.21358 &  0.00005 &  0.06 & 200 &   5129.40036   &  Field    \\
V29 &  130 & 05:35:47.54 &  +34:12:14.9 & 13.241 &  0.653 &   0.848 &  0.21400 &  0.00005 &  0.02 & 170 &   5134.28896   &  Field    \\
V30 &  931 & 05:36:09.86 &  +34:07:06.7 & 17.046 &  1.204 &   1.556 &  0.23585 &  0.00021 &  0.06 & 196 &   5129.24659   &  Field    \\
V31 &  915 & 05:36:42.60 &  +34:10:54.8 & 17.008 &  1.256 &   1.459 &  0.23663 &  0.00006 &  0.07 & 154 &   5129.23537   &  Field    \\
V32 &  972 & 05:36:28.26 &  +34:08:10.3 & 17.133 &  1.250 &   1.629 &  0.23861 &  0.00005 &  0.08 & 193 &   5129.38604   &  Field    \\
V33 &  279 & 05:35:51.49 &  +34:09:27.6 & 14.748 &  0.674 &   0.901 &  0.24272 &  0.00006 &  0.04 & 225 &   5129.32250   &  Field    \\
V34 & 1561 & 05:36:25.83 &  +34:07:33.8 & 18.279 &  1.153 &   1.144 &  0.27196 &  0.00746 &  0.09 & 152 &   5129.28818   &  Field    \\
V35 &  196 & 05:36:17.84 &  +34:09:14.9 & 13.986 &  0.520 &   0.717 &  0.27632 &  0.00008 &  0.03 & 231 &   5129.39834   &  {\bf Cluster}  \\
V36 &  862 & 05:35:48.18 &  +34:09:15.6 & 17.035 &  1.165 &   1.291 &  0.28555 &  0.00008 &  0.09 & 175 &   5134.30780   &  {\bf Cluster}  \\
V37 & 1687 & 05:36:01.54 &  +34:13:49.1 & 18.410 &  1.125 &   1.746 &  0.28944 &  0.00008 &  0.13 & 162 &   5129.17770   &  Field    \\
V38 &  368 & 05:36:48.03 &  +34:12:08.8 & 15.176 &  0.913 &   1.004 &  0.29360 &  0.00009 &  0.09 &  80 &   5129.11522   &  Field    \\
V39 & 1385 & 05:36:10.11 &  +34:11:22.4 & 18.405 &   -    &   1.748 &  0.30441 &  0.00014 &  0.16 & 111 &   5853.42912   &  {\bf Cluster}  \\
V40 &   67 & 05:36:16.64 &  +34:05:01.2 & 12.154 &  0.151 &   0.253 &  0.31143 &  0.00010 &  0.02 & 193 &   5129.40390   &  {\bf Cluster}  \\
V41 &   72 & 05:36:36.45 &  +34:04:17.5 & 12.271 &  0.178 &   0.230 &  0.32206 &  0.00010 &  0.02 & 186 &   5129.11853   &  {\bf Cluster}  \\
V42 & 1214 & 05:36:22.65 &  +34:06:42.0 & 17.745 &  1.295 &   1.803 &  0.36023 &  0.00019 &  0.11 & 180 &   5129.33547   &  {\bf Cluster}  \\
V43 &  110 & 05:36:33.50 &  +34:06:31.8 & 12.937 &  0.785 &   0.877 &  0.36430 &  0.02041 &  0.03 & 213 &   5129.14672   &  Field    \\
V44 & 1699 & 05:36:15.59 &  +34:06:59.2 & 18.499 &   -    &   2.041 &  0.43122 &  0.00020 &  0.21 & 121 &   5144.11932   &  Field    \\
V45 &  787 & 05:35:56.98 &  +34:07:35.8 & 16.808 &  1.148 &   1.363 &  0.43821 &  0.00013 &  0.10 & 111 &   5129.31151   &  Field    \\
V46 & 1209 & 05:36:24.43 &  +34:05:44.8 & 17.769 &   -    &   2.033 &  0.44603 &  0.00020 &  0.15 & 163 &   5129.23164   &  {\bf Cluster}  \\
V47 & 1393 & 05:36:29.95 &  +34:08:38.1 & 18.051 &  1.080 &   1.427 &  0.45310 &  0.00016 &  0.15 & 170 &   5129.16682   &  Field    \\
V48 &  424 & 05:36:17.98 &  +34:05:38.9 & 15.438 &  0.939 &   1.236 &  0.47483 &  0.00023 &  0.08 & 233 &   5129.39834   &  {\bf Cluster}  \\
V49 &  307 & 05:36:10.44 &  +34:08:07.0 & 14.959 &  0.709 &   0.869 &  0.51546 &  0.02345 &  0.03 & 226 &   5129.31266   &  {\bf Cluster}  \\
V50 &  456 & 05:35:48.61 &  +34:13:30.4 & 15.626 &  1.000 &   1.112 &  0.53533 &  0.00029 &  0.05 & 184 &   5134.99268   &  Field    \\
V51 &  702 & 05:36:18.04 &  +34:09:31.0 & 16.414 &  1.069 &   1.398 &  0.55006 &  0.00017 &  0.07 & 208 &   5128.88825   &  Field    \\
V52 &  324 & 05:36:39.98 &  +34:08:57.6 & 15.009 &  0.728 &   0.845 &  0.55157 &  0.00030 &  0.06 & 221 &   5129.01372   &  {\bf Cluster}  \\
V53 &    7 & 05:36:42.30 &  +34:12:06.0 &  9.133 &  0.050 &    -    &  0.59701 &  0.00036 &  0.07 &  45 &   5129.16055   &  {\bf Cluster}  \\
V54 &  600 & 05:36:44.29 &  +34:10:30.9 & 15.911 &  1.042 &   1.496 &  0.61996 &  0.00014 &  0.07 & 198 &   5129.05005   &  {\bf Cluster}  \\
V55 &  268 & 05:35:55.78 &  +34:10:07.7 & 14.729 &  0.674 &   0.872 &  0.63052 &  0.00040 &  0.15 & 218 &   5128.91778   &  Field    \\
V56 & 2133 & 05:36:05.11 &  +34:04:49.1 & 19.079 &  1.425 &   1.582 &  0.75301 &  0.00057 &  0.18 &  46 &   5531.08105   &  Field    \\
V57 &  500 & 05:36:13.94 &  +34:04:54.8 & 15.890 &  1.031 &   1.197 &  0.88339 &  0.00599 &  0.09 & 217 &   5129.20716   &  Field    \\
V58 &   33 & 05:36:15.30 &  +34:07:12.7 & 10.792 &  0.163 &   0.192 &  0.88574 &  0.01984 &  0.04 & 227 &   5129.39834   &  {\bf Cluster}  \\
V59 &  875 & 05:36:36.64 &  +34:06:13.1 & 16.989 &  1.285 &   1.478 &  0.92678 &  0.00086 &  0.09 & 194 &   5128.82089   &  {\bf Cluster}  \\
V60 &  212 & 05:36:01.07 &  +34:07:43.7 & 14.144 &  0.567 &   0.745 &  1.02041 &  0.00104 &  0.06 & 234 &   5128.55232   &  {\bf Cluster}  \\
V61 &  200 & 05:35:50.85 &  +34:06:03.6 & 14.070 &  0.672 &   0.900 &  1.07066 &  0.00115 &  0.04 & 224 &   5129.13606   &  Field    \\
V62 &  217 & 05:36:20.04 &  +34:09:14.7 & 14.186 &  0.574 &   0.682 &  1.07066 &  0.00115 &  0.03 & 231 &   5128.57311   &  {\bf Cluster}  \\
V63 &  203 & 05:36:09.47 &  +34:08:51.3 & 14.110 &  0.573 &   0.714 &  1.12867 &  0.00127 &  0.03 & 234 &   5129.00008   &  {\bf Cluster}  \\
V64 &  964 & 05:36:01.98 &  +34:07:36.3 & 17.142 &  0.919 &   1.120 &  1.16782 &  0.00014 &  0.05 & 192 &   5129.25309   &  Field    \\
V65 &  678 & 05:36:21.99 &  +34:10:47.9 & 16.413 &  1.018 &   1.300 &  1.28041 &  0.00164 &  0.11 & 222 &   5128.67677   &  Field    \\
V66 & 1472 & 05:36:08.45 &  +34:11:48.0 & 18.146 &  1.225 &   1.741 &  1.40964 &  0.00020 &  0.17 & 167 &   5128.36648   &  Field    \\
V67 & 1279 & 05:36:28.74 &  +34:07:06.5 & 17.808 &  1.501 &   1.889 &  1.83150 &  0.00034 &  0.10 & 181 &   5129.40036   &  {\bf Cluster}  \\
V68 &  400 & 05:36:31.78 &  +34:04:12.1 & 15.440 &  0.853 &   0.989 &  2.12224 &  0.00045 &  0.07 & 209 &   5127.48751   &  Field    \\
V69 & 1083 & 05:36:45.55 &  +34:14:07.4 & 17.433 &  1.256 &   1.182 &  4.49236 &  0.00202 &  0.10 & 140 &   5126.09630   &  Field    \\
V70 &  330 & 05:35:48.79 &  +34:07:46.7 & 15.019 &  1.012 &   1.144 &  7.75194 &  0.00601 &  0.06 & 212 &   5123.92274   &  Field    \\
V71 &   50 & 05:36:19.48 &  +34:12:14.8 & 11.564 &  0.187 &   0.252 &  8.41120 &  0.00175 &  0.04 & 218 &   5122.12997   &  {\bf Cluster}  \\
V72 & 1513 & 05:36:07.31 &  +34:09:41.6 & 18.172 &  1.324 &   1.915 & 10.74114 &  0.01154 &  0.19 & 168 &   5126.82180   &  Field    \\
\hline
\end{tabular}
\end{table*}
%{*}{These stars are identified as cluster members in the present study.}

%
%----------------------Fig: Periodic Variables  ----------------------
\begin{figure*}
\centering
\centerline{\includegraphics[width=15.0cm,height=11.0cm]{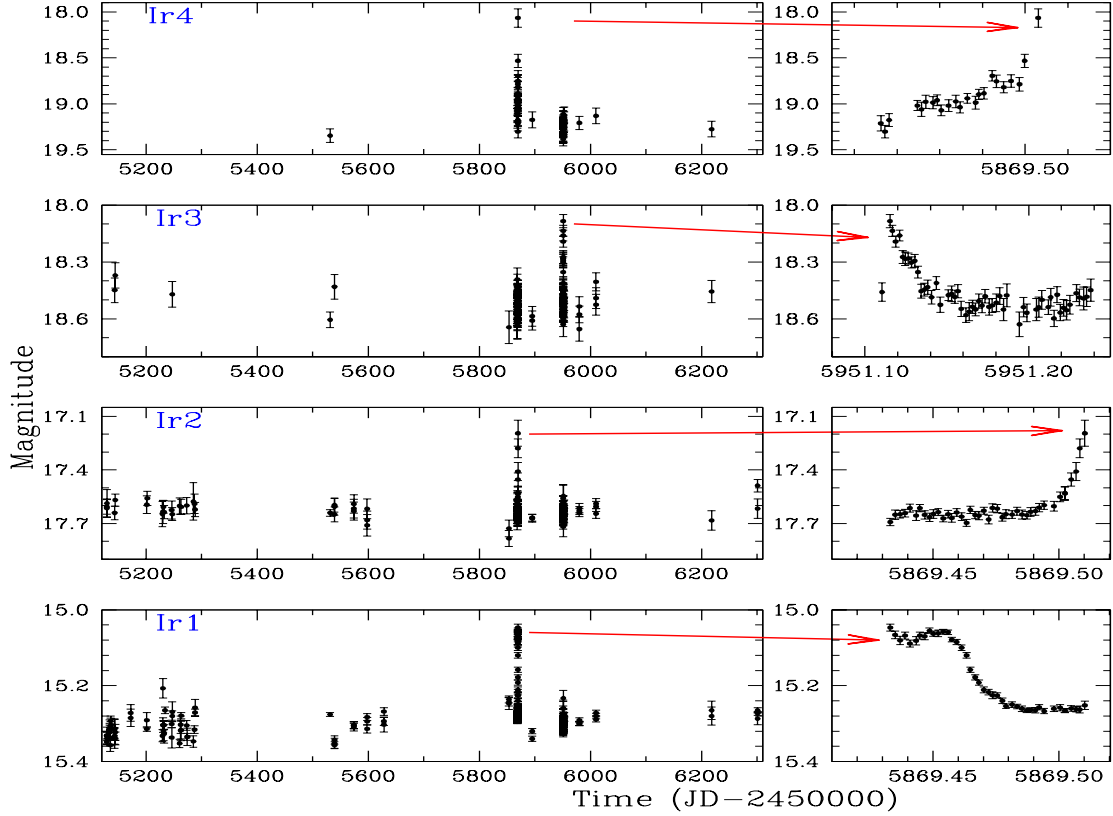}}
%\vspace{-2.5cm}
\caption{The time-magnitude variation of four irregular variables found in this study.}
\label{irregular}
\end{figure*}
%----------------------------- End Fig ----------------------
%
% List of irregular variables
\begin{table*}
\caption{The parameters of four irregular variable stars identified in the cluster NGC~1960. Variables are sorted in increasing magnitude. Here, $\Delta V$ represents the total magnitude variation between the minimum and maximum brightness.}
\centering
  \label{table_irregular}
  \begin{tabular}{cccccccccc}
  \hline
No.  &  ID  &  RA (J2000)  & DEC (J2000) &   V    & (B-V) & (V-I) &$\Delta$ V& N     & Membership\\ 
     &      & (hh:mm:ss)   & (dd:mm:ss)  & (mag)  & (mag) & (mag) &  (mag)   &       &           \\
\hline
Ir1  & 0384 & 05:36:09.40 & +34:07:25.3  & 15.301 & 0.849 & 0.958 &  0.119   &   237 &  {\bf Cluster}  \\
Ir2  & 1169 & 05:36:39.85 & +34:08:19.4  & 17.640 & 1.160 & 1.381 &  0.058   &   183 &  Field    \\ 
Ir3  & 1810 & 05:36:28.14 & +34:06:56.7  & 18.620 & 1.572 & 1.950 &  0.298   &   143 &  Field    \\
Ir4  & 2268 & 05:36:17.19 & +34:05:42.3  & 19.101 & 1.791 & 2.159 &  0.090   &    67 &  {\bf Cluster}  \\
\hline
\end{tabular}
\end{table*}

After our final analysis of light curves, we selected a total of 72 stars in our target field as periodic variables having a wide range in brightness from $V$=9.1 mag to 19.4 mag. They have period ranging from 41 minute to 10.74 day. A binned phase folded light curve was constructed by producing an average of the magnitude and error from the multiple data points within 0.02 phase bins. In Figure~\ref{periodic}, we draw phase folded light curves to illustrate the periodic nature of these 72 variables. In some of the variables, occasional points are scattered away from the periodic cycle which could be due to poor observing conditions. It is evident from the figure that most of these variables show low-amplitude periodicity of the level of few tens of milli-mag. The main characteristics of the variable stars are listed in Table~\ref{table_periodic} which gives identification number, their celestial coordinates, period and type of variability. We give their intensity averaged mean magnitude and amplitude of brightness variation in $V$ band. The variables are arranged in order of increasing period. Since $B$ magnitude of all the stars could not estimated in the present photometry, $(B-V)$ colour of some of the variables could not be ascertained.
%
%-------------------- Fig Finding chart for variables------------
\begin{figure*}
\hbox{
  \hspace{-0.3 cm}
  \includegraphics[width=11.8 cm, height=13.5 cm]{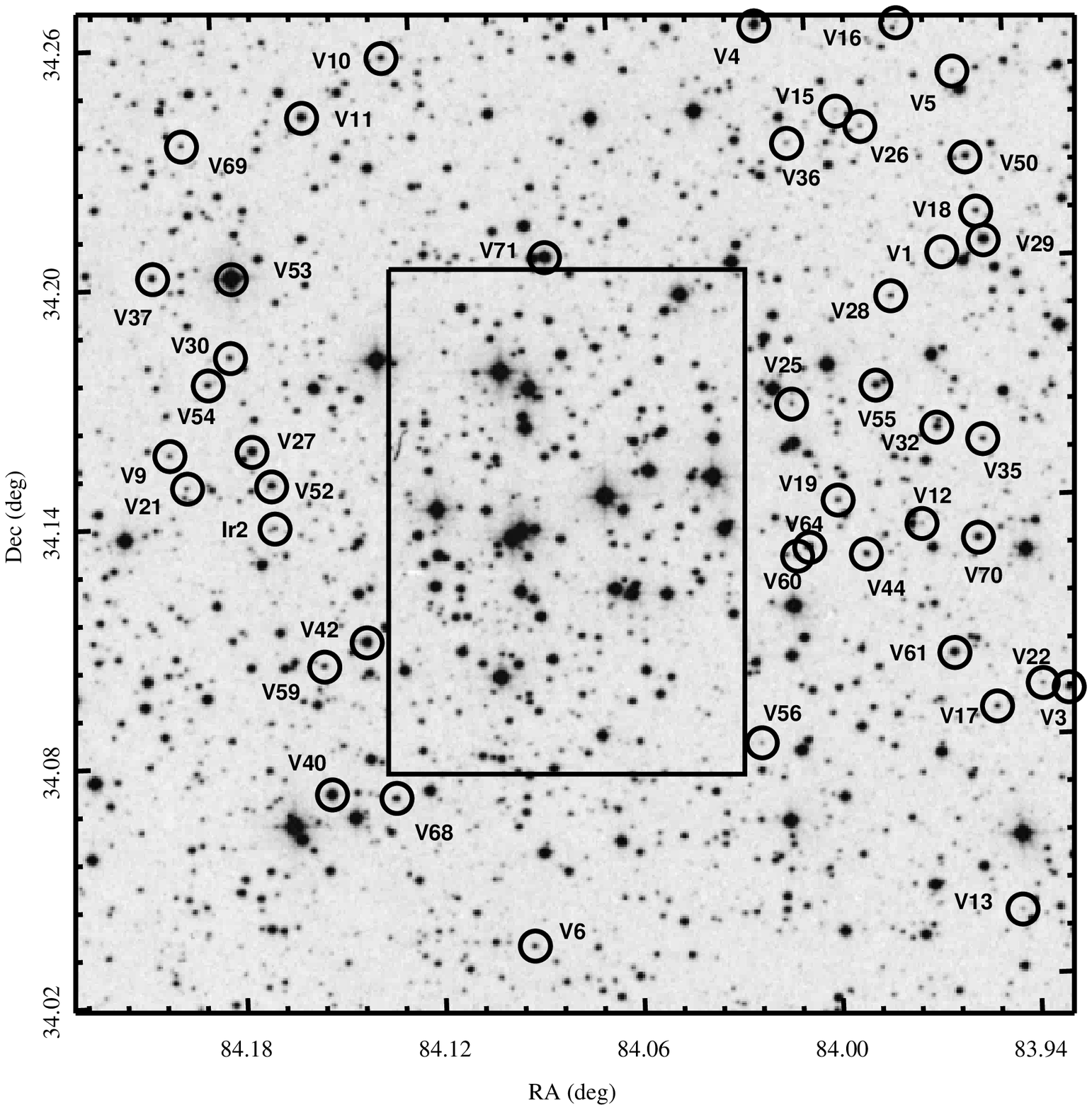}
 \hspace{-0.5 cm}
  \includegraphics[width=7.0 cm, height=12.0 cm]{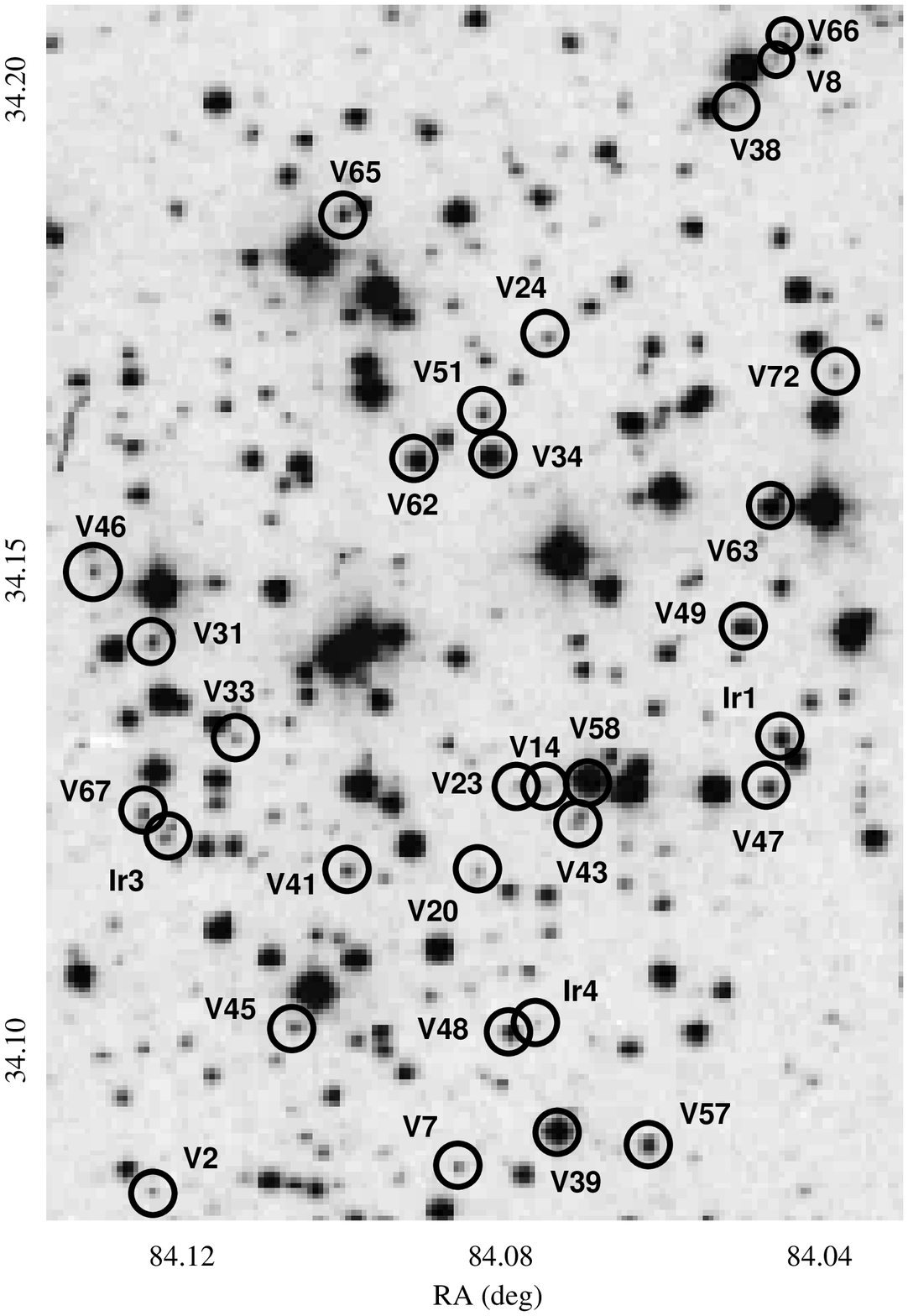}
  }
\caption{The finding chart for the 76 variable stars identified in the field of NGC\,1960. The central region of the cluster marked by rectangular area is shown in the right side of the figure. The positions of the identified variables are marked by circles along with their respective IDs as given in Tables~\ref{table_periodic} and \ref{table_irregular}.}
\label{fchart}
\end{figure*}
%-----------------------------------------------------------
%
%---------------------- Fig. H-R diagram for variables -------------
\begin{figure}
\includegraphics[width=8.5cm,height=6.5cm]{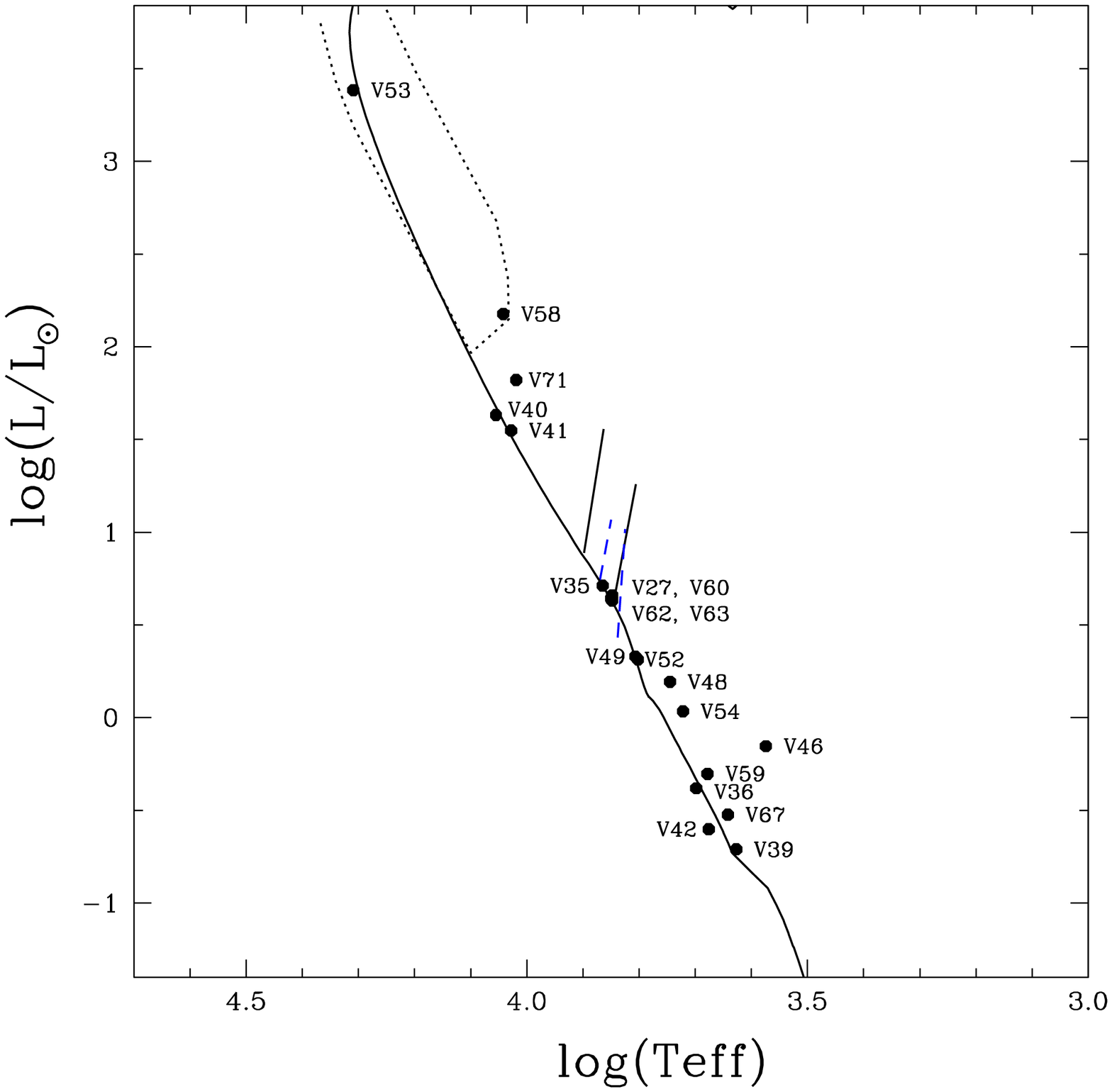}
\vspace{-0.4cm}
\caption{The positions of 20 periodic variables which belong to the cluster NGC\,1960 are shown in the Temperature-Luminosity plane (H-R diagram). The continuous line represents theoretical main-sequence for the $\log$(Age)=7.44$\pm$0.02.The continuous and dashed lines represent regions of instability strips containing the $\delta$-Scuti and $\gamma$-Doradus stars, respectively while the region surrounded by dotted points are locations of slowly pulsating B type stars. All the variable star are marked with their respective IDs.}
\label{hrdia_vari}
\end{figure}
%----------------------------------------------------------
%

\subsection{Irregular variables} \label{iv}
Some stars found in the cluster field seem to show irregular brightness variation in our observations. On careful inspection, we found variability in  four such stars for which we could not determine the period. Either they are long-period variables or their periodic nature could not be ascertained unambiguously due to our intermittent observations. Such variables are assigned as irregular variables. Among them, only two stars are classified as the cluster members. In Figure~\ref{irregular}, we show time-magnitude diagrams of these four stars. Their positions, $V$ magnitude, $(B-V)$ and $(V-I)$ colours along with amplitude in $V$ band are listed in Table~\ref{table_irregular}. These stars show a variation of brightness between 0.3 mag to 1.3 mag during a short span and could be eruptive variables which lasts from tens of seconds to tens of minutes, and then returns to its normal level of brightness on timescales of tens of minutes or hours \citep[e.g.,][]{2015ApJ...814...35C}. On the other hand, some of the variations could be due to multi-periodicity pulsation in the star. We need further observations of these kind of variables over some period of time to understand about their physical nature.

In Figure~\ref{fchart}, we provide a finding  chart of a $\sim 13^\prime \times 13^\prime$ $V$ band CCD frame obtained in our observations, wherein the locations of 76 variable stars identified in the present study are marked by the circles. This manifest that most of the variables are found to be located in the central region of the cluster.
\section{Characterization of variable stars} \label{charac}
Barring one, all the variable stars identified in the present study are newly discovered variables. Therefore, we first determine their physical parameters before characterize their nature.
\subsection{Physical parameters of cluster variables} \label{parameter}
To check whether variable stars identified in the present study are associated with the cluster, we cross-checked them in the list of cluster members given in Table~\ref{mpm}. Out of the 76 variables identified in the present study, only 22 of them belong to the cluster and remaining 54 belong to field stars population lying in the direction of cluster region. Among 22 variables found in the cluster, 20 are periodic variables. As NGC\,1960 is a populous cluster field in the Galactic plane, it is not surprising that many variables detected by us are actually field stars.

As all the 20 variables fall in the main-sequence of the cluster, we determined their physical parameters using the well known relations. The intrinsic magnitude and color of the cluster variables are determined using the distance modulus $(m-M)_0$ = 10.33 mag and extinction $E(B-V)$ = 0.24 mag as estimated in Sect.~\ref{para}. The effective temperature $T_{eff}$ of the star was determined from $(B-V)_0$ using the relation given by \citet{2010AJ....140.1158T}. For two stars, $(B-V)_0$  was not available so we transformed $(V-I)_0$ to $(B-V)_0$ using the standard colour equation. We estimated bolometric magnitude, $M_{bol}$, of each star using the relation $M_{bol}= M_V + BC$ where $BC$ is the bolometric correction that was estimated using the $T_{eff}$ . The luminosity of the variable stars were estimated using the relation $log\,(L/L_\odot) = -0.4\,(M_{bol} - M_{bol_\odot})$ where $M_{bol_\odot}$ is the bolometric magnitude of the Sun which was taken as 4.73 mag \citep{2010AJ....140.1158T}. The parameters, luminosity [$log\,(L/L_\odot$)], bolometric magnitude ($M_{bol}$), effective temperature ($T_{eff}$) and bolometric correction ($BC$), estimated for the 20 periodic variables assigned as cluster members are listed in Table~\ref{MS_parameters}. Stellar masses of these periodic variables have been determined by placing each object on the colour-magnitude diagram and comparing their positions with the mass tracks of Marigo's theoretical isochrones for solar metallicity \citep{2017ApJ...835...77M}. In Table~\ref{MS_parameters}, we give derived masses of these stars where most massive variable star in the cluster is found to have a mass of $\sim 7 M_\odot$.
\subsection{H-R diagram and classification of the cluster variables} \label{hrdia}
The CMD is very useful in separating different class of variable stars. In Figure~\ref{hrdia_vari}, we show the position of these 20 periodic variables belonging to the cluster in the temperature-luminosity H-R diagram. Here, we also draw theoretical isochrones of \citet{2017ApJ...835...77M} as discussed in Sect.~\ref{cmd}. If we examine the locations of these 20 variable stars in the H-R diagram of the cluster, we found that all of them nicely fall along the main-sequence except two or three stars which seems to be reddened cluster members. In Figure~\ref{hrdia_vari}, we also show the positions of various instability strips in the H-R diagram. The theoretical instability strips for $\delta$-Scuti, $\gamma$-Doradus stars and slowly pulsating B stars (SPBs) are shown by continuous, dashed, and dotted lines, respectively taken from \citet{2011MNRAS.413.2403B}, and references therein.

In order to understand nature of these 20 periodic stars, we assessed the classifications of variable stars by comparing their phase-folded light curves to the  exemplar light curves for different class of variable stars. We examined the nature of their variability based on the primary observational properties such as shape of their light curves, periodicity, broad magnitude ranges, spectral classes, amplitudes of the variability and locations in the H-R diagram to the extent possible. When we inspect positions of 20 cluster variables within various instability strips in the H-R diagram as drawn in Figure~\ref{hrdia_vari} as well their distinguish characteristics, we classify (i) 2 $\delta$-Scuti stars, (ii) 3 $\gamma$-Doradus stars, (iii) 2 SPB stars, (iv) 5 rotational variables, (v) 2 non-pulsating stars, and rest 6 stars as miscellaneous class of variables which do not fall in any particular category. In the following subsections, we individually describe the nature of each variability class. 
\subsubsection{$\delta$-Scuti variables} \label{delta-scuti}
$\delta$-Scuti stars are $p$-mode pulsating variables with period smaller than 0.3 day. They belong to spectral type between $A2$ to $F2$ and locate within the $\delta$-Scuti instability strips. We found two stars, V27 and V35 in the cluster whose characteristics are similar to $\delta$-Scuti variables. The variable V35 is located in the middle of the blue and red edges of $\delta$-Scuti instability strip. However, V27 is close to the cool border of $\delta$-Scuti instability strip which also overlaps with the $\gamma$-Doradus instability strip. However, it has very low period of about 0.208 d which makes it an ideal candidate for the $\delta$-Scuti pulsator. However, one cannot rule it out V27 as a possible $\delta$~Scuti-$\gamma$-Doradus hybrid variable as many such hybrid candidates are already known in the open clusters \citep[e.g.,][]{2008ApJ...675.1254H, 2012MNRAS.419.2379J} and they offer vital constraints on the stellar structure due to their simultaneous existence of two different pulsations modes.
\subsubsection{$\gamma$-Doradus variables} \label{gamma-dor}
$\gamma$-Doradus are multi-periodic variable star pulsating in the $g$-modes. These are typically young, early $F$ or late $A$ type main-sequence stars with periods in the range of about 0.3 to 3 day and brightness fluctuations $\sim$ 0.1 mag \citep{2011MNRAS.415.3531B}. They fall in a fairly small region in the $\gamma$-Doradus instability strips which is typically below the $\delta$-Scuti instability strip though some portion of the instability strips of these two classes overlaps. On the basis of physical characteristics of 20 cluster variables and their position in the H-R diagram, we found three stars V60, V62 and V63 which belong to the class of $\gamma$-Doradus variables. Though these three stars lie very close to the blue edge of the $\delta$-Scuti instability strip, but their high period in excess of 0.56 day suggest that they could be $\gamma$-Doradus star.
\subsubsection{Slowly pulsating B type stars} \label{spb}
SPB stars are pulsating stars having periods 0.5 day up to a few days. Their instability strip in the H-R diagram is shown by the dotted line in Figure~\ref{hrdia_vari}. These are two bright stars, V53 and V58, having brightness more than 10.8 mag and temperature larger than 10000 K falling within the region of SPB instability strips and we classified them as SPB stars.

The brightest star in our catalogue of variables is $V53$ which is a well known bright star BD +34 1113 in the SIMBAD and reported to be an eruptive variable in the GCVS catalogue. In our study, the bolometric magnitude of this star comes out to be about -3.73 mag with an effective temperature of $\sim$ 20,000 K which confirms its designated spectral type by of $B2Ve$ by \citet{2012MNRAS.420.2884S} who found very high-level of IR-excess in this star. $V53$ is reported to be a variable in the Hipparcos data with a period of 16.86 day in the AAVSO catalogue. Though star got saturated in most of our frames but on the basis of 43 data points where magnitude of this star could be determined, we found a periodicity of 0.597 day having an amplitude of 0.06 mag. Its small period, high temperature and location in the SPB instability strips suggests that this star is indeed a SPB star.
\subsubsection{Rotational variables} \label{rot}
Stellar rotation and magnetic activity are normally associated with a main-sequence star of $G$ or later spectral type. These stars are characterized by small amplitude, typically less than 0.1 mag and red in colour $(B-V)_0 > 0.5$ mag.  To identify the rotational variables in the present study, we first identify those cluster members which have late-type spectral class. Among 20 cluster members which show periodicity in the present study, 5 stars are found to have $(B-V)_0$ redder than 0.5 mag and show less than 0.10 mag brightness variations in $V$ band. We characterize them as rotational variables. It is very much possible that periodic variation of these 5 stars are due to cool photospheric spots whose brightness is modulated by the stellar rotation. The number of cool spots in these stars is related to their surface magnetic field. It is reported that the level of magnetic activity is larger in young clusters \citep{2010A&A...513A..29M}. As NGC\,1960 is a young cluster having an age of about 27 Myr, it is not surprising that many stars in the cluster are found to be rotational variables.
%
% Characterstics of cluster variables
\begin{table}
\caption{The basic parameters estimated for the 20 periodic variables found in the cluster NGC~1960. The classification of the variables on the basis of their characteristics is given in the last column.}
\label{MS_parameters}
\small
\begin{tabular}{llccccc}
\hline
   ID    & $\log T_{\rm eff}$ &  $BC_V$ & $M_{bol}$ & $\log(L/L_{\odot})$&   Mass       & Type \\
         &      (dex, K)      &  (mag)  &  (mag)    &        (dex)       & ($M_\odot$)  &      \\
\hline                                                                                 
    V27  &     3.843          &   0.062 &   2.995   &  0.694             &    1.53      &  $\delta$-Sct   \\  
    V35  &     3.865          &   0.039 &   2.951   &  0.712             &    1.53      &  $\delta$-Sct  \\  
    V36  &     3.698          &  -0.277 &   5.684   & -0.381             &    0.90      &  Rotational   \\  
    V39  &     3.627          &  -0.822 &   6.509   & -0.711             &    0.80      &  Misc   \\  
    V40  &     4.055          &  -0.430 &   0.650   &  1.632             &    2.69      &  Non-pulsating \\
    V41  &     4.028          &  -0.336 &   0.861   &  1.548             &    2.58      &  Non-pulsating \\
    V42  &     3.676          &  -0.436 &   6.235   & -0.602             &    0.83      &  Misc   \\  
    V46  &     3.574          &  -1.580 &   5.115   & -0.154             &    0.83      &  Misc   \\  
    V48  &     3.745          &  -0.117 &   4.247   &  0.193             &    1.17      &  Rotational   \\  
    V49  &     3.807          &   0.020 &   3.905   &  0.330             &    1.28      &  Misc   \\  
    V52  &     3.802          &   0.012 &   3.947   &  0.313             &    1.26      &  Misc  \\  
    V53  &     4.309          &  -1.797 &  -3.731   &  3.384             &    7.10      &  SPB    \\  
    V54  &     3.722          &  -0.191 &   4.646   &  0.034             &    0.98      &  Rotational   \\  
    V58  &     4.042          &  -0.430 &  -0.712   &  2.177             &    4.40      &  SPB   \\  
    V59  &     3.678          &  -0.428 &   5.487   & -0.303             &    0.90      &  Rotational   \\  
    V60  &     3.850          &   0.055 &   3.125   &  0.642             &    1.49      &  $\gamma$-Dor   \\  
    V62  &     3.848          &   0.039 &   3.151   &  0.632             &    1.47      &  $\gamma$-Dor   \\  
    V63  &     3.848          &   0.043 &   3.079   &  0.660             &    1.50      &  $\gamma$-Dor   \\  
    V67  &     3.642          &  -0.693 &   6.041   & -0.524             &    0.82      &  Rotational   \\  
    V71  &     4.019          &  -0.312 &   0.177   &  1.821             &    3.34      &  Misc   \\  
\hline
\end {tabular}
\end{table}

\subsubsection{Non-pulsating variables} \label{npul}
We found two bright $B$ type stars V40 and V41 between the cool edge of the SPB and the hot edge of the $\delta$-Scuti instability strips and found to have similar properties like period ($\sim$ 0.3 day), amplitude ($\sim$ 0.02 mag) and mass ($\sim$ 2.6 M$_\odot$). We classify these stars as non-pulsating variables as suggested by \citet{2011MNRAS.413.2403B}. The origin of this grouping as non-pulsating variable is not clear, but may be related to the rotation. On the basis of such stars found between the $\beta$-Cep and SPB instability strips in the Kepler data, \citet{2011MNRAS.413.2403B} suggested some of them may be binary stars. In young open clusters NGC\,3766 and Stock\,8, \citet{2013A&A...554A.108M} and \citet{2019AJ....158...68L} have also found a large population of new variables between SPB and $\delta$-Scuti stars, the region where no pulsation mechanisms were expected on the basis of theoretical evolutionary models.
\subsubsection{Miscellaneous variables} \label{misc}
There are some periodic stars which could not be classified in any particular class of variables on the basis of their estimated parameters and phased light curves. We found six such variables in the cluster namely V39, V42, V46, V49, V52 and V71 which are classified as Miscellaneous variables in the present study and marked as $Misc$ in Table~\ref{MS_parameters}. While first five stars are relatively faint, the variable V71 is quite a bright star lying in the region between the instability strips of the $\delta$-Scuti and SPB stars.
\subsection{Eclipsing binaries} \label{eb}
On the basis of phased light curves of 72 periodic variables, we identified only star having ID V47 as eclipsing binary system (EBs). This star with a rotation period of 0.4717 day clearly shows two eclipses with different depth. Classical approach of frequency analysis does not succeed to extract true period of EBs, so we estimated period of V47 from the eclipse minima and by the visual inspection of the phase diagram for the multiple periods. According to the variability type listed in the Moscow General Catalog of Variable Stars (GCVS), we classified V47 as W UMa eclipsing binary star. The star is found to belong field population according to its membership probability ($p$ = 0.01) given in Table~\ref{catalog}. As it is not possible to do a detailed analysis of V47 in the present analysis, a follow-up paper is in preparation that includes modelling of its photometric light curves and spectroscopic analysis in order to determine its physical parameters and examination of stellar spots. It should be noted that many EBs might have been escaped from our detection due to short time base and low duty cycle, particularly when most of the photometric variations we detected are extremely low amplitudes.
\section{Discussion and Conclusions}\label{summary}
The cluster parameter based on the photometric analysis may be subject to selection bias as many genuine cluster members are left out from the sample. Though lack of the identification of true members may not have much effect on the determinations of parameters like reddening, age, distance modulus and metallicity but imparts significant impact on the dynamical study of the cluster like mass segregation, relaxation time, and half cluster radius. It is however extremely difficult to extract all the genuine members of a cluster on the basis of photometric observations due to their identification issue. On the other hand, cluster membership assignments based on the kinematic studies have always been considered more reliable than those obtained through the photometric analysis.

In the present study, we performed an extensive photometric and kinematic investigation of a relatively young open cluster NGC\,1960. We made a multi-band photometric catalogue of 3962 stars in the cluster by supplementing our $UBVRI$ data with the SHA06 photometric data along with the archival photometric, near-IR and kinematic data.  We determined membership probabilities of the stars based on their kinematic data provided by the most accurate proper motions catalogue produced till date through the Gaia DR2 survey. As membership probabilities cannot be sole criteria to identify the true cluster members, we further used stellar parallaxes to isolate cluster members. The precision of the Gaia DR2 proper motions coupled with the strength to distinguish cluster members through its parallax measurements allowed us to isolated cluster members quite remarkably. We found a total of 262 stars which belong to the cluster NGC\,1960. This number is relatively low in comparison of the total number of stars found in the target field. However, it is not surprising as the cluster lies very close to the Galactic mid-plane ($b \approx 1$ deg) due to which Galactic field star population is very dominant in the target field. We obtained a mean cluster parallax of 0.86$\pm$0.05 mas, excluding the stars showing large errors ($e\overline\omega>0.2$) in their parallax measurements. This corresponds to a mean distance of $\sim$ 1.17$\pm$0.06 kpc and distance modulus of $(m-M)_0 = 10.33\pm0.11$ mag. The mean proper motion of the cluster was determined to be -0.143$\pm$0.008~mas/yr and -3.395$\pm$0.008~mas/yr in the direction of RA and DEC, respectively. On the basis of $(U-B)/(B-V)$ colour-colour diagram, the reddening $E(B-V)$ was estimated as 0.24$\pm$0.02 mag in the optical bands which was found to be 0.23 mag in the near-IR data, in agreement with the optical reddening. Since NGC\,1960 shows some signature of differential extinction across the cluster region in its $(U-B)/(B-V)$ diagram, it indicates that the cluster may still be embedded within the parent molecular cloud. We estimated an average total-to-selective extinction ratio as 3.10$\pm$0.08 that is in excellent agreement to the normal value. However, colour-excess ratio $E(U-B)/E(B-V)$ is found to be slightly higher than the normal one. Our measurement of reddening gives a visual extinction of $A_V = 0.74\pm0.08$ mag in the direction of the cluster. Exploiting prior knowledge of reddening through colour-colour diagram, and distance through the parallaxes of cluster members, we determined the age of the cluster through the colour-magnitude diagrams. We obtained an age of $27.5^{+1.3}_{-1.2}$ Myr for the cluster NGC\,1960 by visually fitting a recently available solar metallicity isochrones of \citet{2017ApJ...835...77M}.

Since our observations are complete up to 19 mag, we constructed the luminosity function up to this brightness limit only which then converted into the mass function. The mass function slope (MF slope) in the cluster was determined for the stars in the mass range  $0.72 \le M/M_{\odot} \le 7.32$ and MF slope was found to be $\Gamma = -1.26\pm0.19$ which is nearly equal to the Salpeter value of $\Gamma = -1.35$ in the solar neighbourhood. This is well expected considering the cluster NGC\,1960 is relatively young and dynamical evolution has not changed the primordial MF in a significant way. Our result of MF slope consistent with Salpeter value further validates the universal IMF in the Milky way, even though star clusters have a wide range of properties across the Galaxy \citep[e.g.,][]{2019arXiv191107267C}. We also constructed mass function slopes for the inner and outer regions, and observed that the slope is flatter in the inner region than in the outer region, suggesting an ongoing mass segregation process in the cluster. The relaxation time of the cluster was found to be smaller than its age which implies that the cluster is not yet dynamically relaxed. Using the 262 members stars for which $V$ magnitudes are available down to 20.7 mag, we derived a total cluster mass of $\approx$ 417 M$_\odot$ with a mean stellar mass of $\approx$ 1.6 M$_\odot$. Although our selection criteria retrieves most of the cluster members, we still emphasize that the estimated cluster mass could be a lower-limit to the actual total mass of the cluster and mean mass of 1.6 M$_{\odot}$ be considered as upper limit of the stellar mass for the same reason.

The photometric and kinematic studies of NGC\,1960 have been done extensively in the past, however, no variability study was performed on this cluster so far. As search for variable stars is one of the primary goal of our ongoing survey, we carried out a long-term observations of the cluster NGC\,1960 in the $V$ band. We monitored the central $13^{\prime}\times13^{\prime}$ region of NGC\,1960 on 43 nights over a period of more than three years. As we found out 1386 stars in the target field, the search for the variability among these stars has been performed. Through the present survey, we have first provided time-series $V$ band photometric analysis of 76 variable stars, all of them newly detected except one. The variables range in $V$-band magnitudes from 9.1 mag to 19.4 mag. Among 76 stars detected in the present study, 72 stars are found to be  periodic variables with a period range of 41 minutes to 10.74 days. Majority of these stars are short-period variables having period smaller than 1 day. We could not detect any variable with period longer than 10.74 days because of the large gaps between the observing cycles. Most of the short period variables have relatively small amplitudes and we could retrieve amplitude of light variability down to the 0.02 mag level. Out of 72 periodic variables, 20 are identified as cluster members therefore we could obtain their masses, effective temperatures, and bolometric luminosities. Two of the cluster members show irregular variability similar to stellar flares but data is insufficient to visualize any conclusive characteristics. Rest of the 54 variables are identified as field stars population lying in the direction of the cluster. The light curves analysis of 20 cluster variables along with the estimated characteristic parameters suggested that 2 of them may belong to $\delta$-Scuti stars, 3 could be $\gamma-$Doradus type stars, 5 as rotational variables, 2 as SPB stars, and 2 non-pulsating B stars. We could not classify 6 stars in any specific category and characterized them as miscellaneous variables. We also found one star V47 as W UMa eclipsing binary star belonging to the field stars population. We compared our catalog of variable stars with the variables listed in the AAVSO International Variable Star Catalogue \citep{2017yCat....102027W} and found only one common variable which is identified as BD+34 1113 in the SIMBAD. On the basis of Gaia DR2 kinematic data, this star belong to the cluster NGC\,1960 and we classified it as a slowly pulsating B type star in the present study.
\section*{Acknowledgments}
We are thankful to various observers of 104-cm ST for their contributions in accumulating photometric data of this cluster during 2009-2013. We are also grateful to Saurabh Sharma for providing photometric catalogue that has been used in the present study. We used data from the Two Micron All Sky Survey, which is a joint project of the University of Massachusetts; the Infrared Processing and Analysis Center/California Institute of Technology, funded by the NASA. This work has made use of data from the European Space Agency (ESA) mission Gaia (https://www.cosmos.esa.int/Gaia), processed by the Gaia Data Processing and Analysis Consortium (DPAC, https://www.cosmos.esa.int/web/GAIA/dpac/consortium).

\bibliographystyle{mnras}
\bibliography{joshi}

\begin{thebibliography}{}
\makeatletter
\relax
\def\mn@urlcharsother{\let\do\@makeother \do\$\do\&\do\#\do\^\do\_\do\%\do\~}
\def\mn@doi{\begingroup\mn@urlcharsother \@ifnextchar [ {\mn@doi@}
  {\mn@doi@[]}}
\def\mn@doi@[#1]#2{\def\@tempa{#1}\ifx\@tempa\@empty \href
  {http://dx.doi.org/#2} {doi:#2}\else \href {http://dx.doi.org/#2} {#1}\fi
  \endgroup}
\def\mn@eprint#1#2{\mn@eprint@#1:#2::\@nil}
\def\mn@eprint@arXiv#1{\href {http://arxiv.org/abs/#1} {{\tt arXiv:#1}}}
\def\mn@eprint@dblp#1{\href {http://dblp.uni-trier.de/rec/bibtex/#1.xml}
  {dblp:#1}}
\def\mn@eprint@#1:#2:#3:#4\@nil{\def\@tempa {#1}\def\@tempb {#2}\def\@tempc
  {#3}\ifx \@tempc \@empty \let \@tempc \@tempb \let \@tempb \@tempa \fi \ifx
  \@tempb \@empty \def\@tempb {arXiv}\fi \@ifundefined
  {mn@eprint@\@tempb}{\@tempb:\@tempc}{\expandafter \expandafter \csname
  mn@eprint@\@tempb\endcsname \expandafter{\@tempc}}}

\bibitem[\protect\citeauthoryear{{Allison}}{{Allison}}{2012}]{2012MNRAS.421.3338A}
{Allison} R.~J.,  2012, \mn@doi [Monthly Notices of the Royal Astronomical
  Society] {10.1111/j.1365-2966.2012.20557.x}, \href
  {https://ui.adsabs.harvard.edu/abs/2012MNRAS.421.3338A} {421, 3338}

\bibitem[\protect\citeauthoryear{{Balaguer-N{\'u}nez}, {Tian}  \&
  {Zhao}}{{Balaguer-N{\'u}nez} et~al.}{1998}]{1998A&AS..133..387B}
{Balaguer-N{\'u}nez} L.,  {Tian} K.~P.,   {Zhao} J.~L.,  1998, \mn@doi [\aaps]
  {10.1051/aas:1998324}, \href
  {https://ui.adsabs.harvard.edu/abs/1998A&AS..133..387B} {133, 387}

\bibitem[\protect\citeauthoryear{{Balona} et~al.,}{{Balona}
  et~al.}{2011a}]{2011MNRAS.413.2403B}
{Balona} L.~A.,  et~al., 2011a, \mn@doi [\mnras]
  {10.1111/j.1365-2966.2011.18311.x}, \href
  {https://ui.adsabs.harvard.edu/abs/2011MNRAS.413.2403B} {413, 2403}

\bibitem[\protect\citeauthoryear{{Balona}, {Guzik}, {Uytterhoeven}, {Smith},
  {Tenenbaum}  \& {Twicken}}{{Balona} et~al.}{2011b}]{2011MNRAS.415.3531B}
{Balona} L.~A.,  {Guzik} J.~A.,  {Uytterhoeven} K.,  {Smith} J.~C.,
  {Tenenbaum} P.,   {Twicken} J.~D.,  2011b, \mn@doi [\mnras]
  {10.1111/j.1365-2966.2011.18973.x}, \href
  {https://ui.adsabs.harvard.edu/abs/2011MNRAS.415.3531B} {415, 3531}

\bibitem[\protect\citeauthoryear{{Balona}, {Joshi}, {Joshi}  \&
  {Sagar}}{{Balona} et~al.}{2013}]{2013MNRAS.429.1466B}
{Balona} L.~A.,  {Joshi} S.,  {Joshi} Y.~C.,   {Sagar} R.,  2013, \mn@doi
  [\mnras] {10.1093/mnras/sts429}, \href
  {https://ui.adsabs.harvard.edu/abs/2013MNRAS.429.1466B} {429, 1466}

\bibitem[\protect\citeauthoryear{{Barkhatova}, {Zakharova}, {Shashkina}  \&
  {Orekhova}}{{Barkhatova} et~al.}{1985}]{1985AZh....62..854B}
{Barkhatova} K.~A.,  {Zakharova} P.~E.,  {Shashkina} L.~P.,   {Orekhova} L.~K.,
   1985, \azh, \href {https://ui.adsabs.harvard.edu/abs/1985AZh....62..854B}
  {62, 854}

\bibitem[\protect\citeauthoryear{{Bell}, {Naylor}, {Mayne}, {Jeffries}  \&
  {Littlefair}}{{Bell} et~al.}{2013}]{2013MNRAS.434..806B}
{Bell} C. P.~M.,  {Naylor} T.,  {Mayne} N.~J.,  {Jeffries} R.~D.,
  {Littlefair} S.~P.,  2013, \mn@doi [\mnras] {10.1093/mnras/stt1075}, \href
  {https://ui.adsabs.harvard.edu/abs/2013MNRAS.434..806B} {434, 806}

\bibitem[\protect\citeauthoryear{{Camarillo}, {Mathur}, {Mitchell}  \&
  {Ratra}}{{Camarillo} et~al.}{2018}]{2018PASP..130b4101C}
{Camarillo} T.,  {Mathur} V.,  {Mitchell} T.,   {Ratra} B.,  2018, \mn@doi
  [\pasp] {10.1088/1538-3873/aa9b26}, \href
  {https://ui.adsabs.harvard.edu/abs/2018PASP..130b4101C} {130, 024101}

\bibitem[\protect\citeauthoryear{{Cantat-Gaudin} et~al.,}{{Cantat-Gaudin}
  et~al.}{2018}]{2018A&A...618A..93C}
{Cantat-Gaudin} T.,  et~al., 2018, \mn@doi [\aap]
  {10.1051/0004-6361/201833476}, \href
  {https://ui.adsabs.harvard.edu/abs/2018A&A...618A..93C} {618, A93}

\bibitem[\protect\citeauthoryear{{Cardelli}, {Clayton}  \& {Mathis}}{{Cardelli}
  et~al.}{1989}]{1989ApJ...345..245C}
{Cardelli} J.~A.,  {Clayton} G.~C.,   {Mathis} J.~S.,  1989, \mn@doi [\apj]
  {10.1086/167900}, \href
  {https://ui.adsabs.harvard.edu/abs/1989ApJ...345..245C} {345, 245}

\bibitem[\protect\citeauthoryear{{Carpenter}}{{Carpenter}}{2001}]{2001AJ....121.2851C}
{Carpenter} J.~M.,  2001, \mn@doi [\aj] {10.1086/320383}, \href
  {https://ui.adsabs.harvard.edu/abs/2001AJ....121.2851C} {121, 2851}

\bibitem[\protect\citeauthoryear{{Carraro}, {Ng}  \& {Portinari}}{{Carraro}
  et~al.}{1998}]{1998MNRAS.296.1045C}
{Carraro} G.,  {Ng} Y.~K.,   {Portinari} L.,  1998, \mn@doi [\mnras]
  {10.1046/j.1365-8711.1998.01460.x}, \href
  {https://ui.adsabs.harvard.edu/abs/1998MNRAS.296.1045C} {296, 1045}

\bibitem[\protect\citeauthoryear{{Carraro}, {Villanova}, {Demarque}, {Moni
  Bidin}  \& {McSwain}}{{Carraro} et~al.}{2008}]{2008MNRAS.386.1625C}
{Carraro} G.,  {Villanova} S.,  {Demarque} P.,  {Moni Bidin} C.,   {McSwain}
  M.~V.,  2008, \mn@doi [\mnras] {10.1111/j.1365-2966.2008.13143.x}, \href
  {https://ui.adsabs.harvard.edu/abs/2008MNRAS.386.1625C} {386, 1625}

\bibitem[\protect\citeauthoryear{{Chang}, {Byun}  \& {Hartman}}{{Chang}
  et~al.}{2015}]{2015ApJ...814...35C}
{Chang} S.~W.,  {Byun} Y.~I.,   {Hartman} J.~D.,  2015, \mn@doi [\apj]
  {10.1088/0004-637X/814/1/35}, \href
  {https://ui.adsabs.harvard.edu/abs/2015ApJ...814...35C} {814, 35}

\bibitem[\protect\citeauthoryear{{Chen}, {Hou}  \& {Wang}}{{Chen}
  et~al.}{2003}]{2003AJ....125.1397C}
{Chen} L.,  {Hou} J.~L.,   {Wang} J.~J.,  2003, \mn@doi [\aj] {10.1086/367911},
  \href {https://ui.adsabs.harvard.edu/abs/2003AJ....125.1397C} {125, 1397}

\bibitem[\protect\citeauthoryear{{Chini} \& {Wargau}}{{Chini} \&
  {Wargau}}{1990}]{1990A&A...227..213C}
{Chini} R.,  {Wargau} W.~F.,  1990, \aap, \href
  {https://ui.adsabs.harvard.edu/abs/1990A&A...227..213C} {227, 213}

\bibitem[\protect\citeauthoryear{{Chumak}, {Platais}, {McLaughlin},
  {Rastorguev}  \& {Chumak}}{{Chumak} et~al.}{2010}]{2010MNRAS.402.1841C}
{Chumak} Y.~O.,  {Platais} I.,  {McLaughlin} D.~E.,  {Rastorguev} A.~S.,
  {Chumak} O.~V.,  2010, \mn@doi [\mnras] {10.1111/j.1365-2966.2009.16021.x},
  \href {https://ui.adsabs.harvard.edu/abs/2010MNRAS.402.1841C} {402, 1841}

\bibitem[\protect\citeauthoryear{{Colman} \& {Teyssier}}{{Colman} \&
  {Teyssier}}{2019}]{2019arXiv191107267C}
{Colman} T.,  {Teyssier} R.,  2019, arXiv e-prints, \href
  {https://ui.adsabs.harvard.edu/abs/2019arXiv191107267C} {p. arXiv:1911.07267}

\bibitem[\protect\citeauthoryear{{Dahm}}{{Dahm}}{2015}]{2015ApJ...813..108D}
{Dahm} S.~E.,  2015, \mn@doi [\apj] {10.1088/0004-637X/813/2/108}, \href
  {https://ui.adsabs.harvard.edu/abs/2015ApJ...813..108D} {813, 108}

\bibitem[\protect\citeauthoryear{{Dalessandro}, {Miocchi}, {Carraro},
  {J{\'\i}lkov{\'a}}  \& {Moitinho}}{{Dalessandro}
  et~al.}{2015}]{2015MNRAS.449.1811D}
{Dalessandro} E.,  {Miocchi} P.,  {Carraro} G.,  {J{\'\i}lkov{\'a}} L.,
  {Moitinho} A.,  2015, \mn@doi [\mnras] {10.1093/mnras/stv395}, \href
  {https://ui.adsabs.harvard.edu/abs/2015MNRAS.449.1811D} {449, 1811}

\bibitem[\protect\citeauthoryear{{Dar}, {Parihar}, {Saleh}  \& {Malik}}{{Dar}
  et~al.}{2018}]{2018NewA...64...34D}
{Dar} A.~A.,  {Parihar} P.~S.,  {Saleh} P.,   {Malik} M.~A.,  2018, \mn@doi
  [\na] {10.1016/j.newast.2018.04.002}, \href
  {https://ui.adsabs.harvard.edu/abs/2018NewA...64...34D} {64, 34}

\bibitem[\protect\citeauthoryear{{Dias}, {Alessi}, {Moitinho}  \&
  {L{\'e}pine}}{{Dias} et~al.}{2002}]{2002A&A...389..871D}
{Dias} W.~S.,  {Alessi} B.~S.,  {Moitinho} A.,   {L{\'e}pine} J.~R.~D.,  2002,
  \mn@doi [\aap] {10.1051/0004-6361:20020668}, \href
  {https://ui.adsabs.harvard.edu/abs/2002A&A...389..871D} {389, 871}

\bibitem[\protect\citeauthoryear{{Dias}, {Monteiro}, {Caetano}, {L{\'e}pine},
  {Assafin}  \& {Oliveira}}{{Dias} et~al.}{2014}]{2014A&A...564A..79D}
{Dias} W.~S.,  {Monteiro} H.,  {Caetano} T.~C.,  {L{\'e}pine} J.~R.~D.,
  {Assafin} M.,   {Oliveira} A.~F.,  2014, \mn@doi [\aap]
  {10.1051/0004-6361/201323226}, \href
  {https://ui.adsabs.harvard.edu/abs/2014A&A...564A..79D} {564, A79}

\bibitem[\protect\citeauthoryear{{Dotti} \& {Fern{\'a}ndez T{\'\i}o}}{{Dotti}
  \& {Fern{\'a}ndez T{\'\i}o}}{2019}]{2019arXiv191104562D}
{Dotti} G.,  {Fern{\'a}ndez T{\'\i}o} J.~M.,  2019, arXiv e-prints, \href
  {https://ui.adsabs.harvard.edu/abs/2019arXiv191104562D} {p. arXiv:1911.04562}

\bibitem[\protect\citeauthoryear{{Frinchaboy} \& {Majewski}}{{Frinchaboy} \&
  {Majewski}}{2008}]{2008AJ....136..118F}
{Frinchaboy} P.~M.,  {Majewski} S.~R.,  2008, \mn@doi [\aj]
  {10.1088/0004-6256/136/1/118}, \href
  {https://ui.adsabs.harvard.edu/abs/2008AJ....136..118F} {136, 118}

\bibitem[\protect\citeauthoryear{{Gaia Collaboration} et~al.,}{{Gaia
  Collaboration} et~al.}{2018}]{2018A&A...616A...1G}
{Gaia Collaboration} et~al., 2018, \mn@doi [\aap]
  {10.1051/0004-6361/201833051}, \href
  {https://ui.adsabs.harvard.edu/abs/2018A&A...616A...1G} {616, A1}

\bibitem[\protect\citeauthoryear{{Genzel} \& {Townes}}{{Genzel} \&
  {Townes}}{1987}]{1987ARA&A..25..377G}
{Genzel} R.,  {Townes} C.~H.,  1987, \mn@doi [\araa]
  {10.1146/annurev.aa.25.090187.002113}, \href
  {https://ui.adsabs.harvard.edu/abs/1987ARA&A..25..377G} {25, 377}

\bibitem[\protect\citeauthoryear{{Hartman}, {Gaudi}, {Holman}, {McLeod},
  {Stanek}, {Barranco}, {Pinsonneault}  \& {Kalirai}}{{Hartman}
  et~al.}{2008}]{2008ApJ...675.1254H}
{Hartman} J.~D.,  {Gaudi} B.~S.,  {Holman} M.~J.,  {McLeod} B.~A.,  {Stanek}
  K.~Z.,  {Barranco} J.~A.,  {Pinsonneault} M.~H.,   {Kalirai} J.~S.,  2008,
  \mn@doi [\apj] {10.1086/527460}, \href
  {https://ui.adsabs.harvard.edu/abs/2008ApJ...675.1254H} {675, 1254}

\bibitem[\protect\citeauthoryear{{Hasan}, {Kilambi}  \& {Hasan}}{{Hasan}
  et~al.}{2008}]{2008Ap&SS.313..363H}
{Hasan} P.,  {Kilambi} G.~C.,   {Hasan} S.~N.,  2008, \mn@doi [\apss]
  {10.1007/s10509-007-9705-3}, \href
  {https://ui.adsabs.harvard.edu/abs/2008Ap&SS.313..363H} {313, 363}

\bibitem[\protect\citeauthoryear{{Herbst}, {Herbst}, {Grossman}  \&
  {Weinstein}}{{Herbst} et~al.}{1994}]{1994AJ....108.1906H}
{Herbst} W.,  {Herbst} D.~K.,  {Grossman} E.~J.,   {Weinstein} D.,  1994,
  \mn@doi [\aj] {10.1086/117204}, \href
  {https://ui.adsabs.harvard.edu/abs/1994AJ....108.1906H} {108, 1906}

\bibitem[\protect\citeauthoryear{{H{\o}g} et~al.,}{{H{\o}g}
  et~al.}{2000}]{2000A&A...355L..27H}
{H{\o}g} E.,  et~al., 2000, \aap, \href
  {https://ui.adsabs.harvard.edu/abs/2000A&A...355L..27H} {355, L27}

\bibitem[\protect\citeauthoryear{{Honma} et~al.,}{{Honma}
  et~al.}{2012}]{2012PASJ...64..136H}
{Honma} M.,  et~al., 2012, \mn@doi [\pasj] {10.1093/pasj/64.6.136}, \href
  {https://ui.adsabs.harvard.edu/abs/2012PASJ...64..136H} {64, 136}

\bibitem[\protect\citeauthoryear{{Hoyle}, {Shanks}  \& {Tanvir}}{{Hoyle}
  et~al.}{2003}]{2003MNRAS.345..269H}
{Hoyle} F.,  {Shanks} T.,   {Tanvir} N.~R.,  2003, \mn@doi [\mnras]
  {10.1046/j.1365-8711.2003.06939.x}, \href
  {https://ui.adsabs.harvard.edu/abs/2003MNRAS.345..269H} {345, 269}

\bibitem[\protect\citeauthoryear{{Jeffries}, {Naylor}, {Mayne}, {Bell}  \&
  {Littlefair}}{{Jeffries} et~al.}{2013}]{2013MNRAS.434.2438J}
{Jeffries} R.~D.,  {Naylor} T.,  {Mayne} N.~J.,  {Bell} C. P.~M.,
  {Littlefair} S.~P.,  2013, \mn@doi [\mnras] {10.1093/mnras/stt1180}, \href
  {https://ui.adsabs.harvard.edu/abs/2013MNRAS.434.2438J} {434, 2438}

\bibitem[\protect\citeauthoryear{{Johnson} \& {Morgan}}{{Johnson} \&
  {Morgan}}{1953}]{1953ApJ...117..313J}
{Johnson} H.~L.,  {Morgan} W.~W.,  1953, \mn@doi [\apj] {10.1086/145697}, \href
  {https://ui.adsabs.harvard.edu/abs/1953ApJ...117..313J} {117, 313}

\bibitem[\protect\citeauthoryear{{Joshi}}{{Joshi}}{2005}]{2005MNRAS.362.1259J}
{Joshi} Y.~C.,  2005, \mn@doi [\mnras] {10.1111/j.1365-2966.2005.09391.x},
  \href {https://ui.adsabs.harvard.edu/abs/2005MNRAS.362.1259J} {362, 1259}

\bibitem[\protect\citeauthoryear{{Joshi}}{{Joshi}}{2007}]{2007MNRAS.378..768J}
{Joshi} Y.~C.,  2007, \mn@doi [\mnras] {10.1111/j.1365-2966.2007.11831.x},
  \href {https://ui.adsabs.harvard.edu/abs/2007MNRAS.378..768J} {378, 768}

\bibitem[\protect\citeauthoryear{{Joshi}, {Pandey}, {Narasimha}  \&
  {Sagar}}{{Joshi} et~al.}{2005}]{Joshi2005}
{Joshi} Y.~C.,  {Pandey} A.~K.,  {Narasimha} D.,   {Sagar} R.,  2005, \mn@doi
  [\aap] {10.1051/0004-6361:20042357}, \href
  {https://ui.adsabs.harvard.edu/#abs/2005A&A...433..787J} {433, 787}

\bibitem[\protect\citeauthoryear{{Joshi}, {Joshi}, {Kumar}, {Mondal}  \&
  {Balona}}{{Joshi} et~al.}{2012}]{2012MNRAS.419.2379J}
{Joshi} Y.~C.,  {Joshi} S.,  {Kumar} B.,  {Mondal} S.,   {Balona} L.~A.,  2012,
  \mn@doi [\mnras] {10.1111/j.1365-2966.2011.19890.x}, \href
  {https://ui.adsabs.harvard.edu/abs/2012MNRAS.419.2379J} {419, 2379}

\bibitem[\protect\citeauthoryear{{Joshi}, {Balona}, {Joshi}  \&
  {Kumar}}{{Joshi} et~al.}{2014}]{2014MNRAS.437..804J}
{Joshi} Y.~C.,  {Balona} L.~A.,  {Joshi} S.,   {Kumar} B.,  2014, \mn@doi
  [\mnras] {10.1093/mnras/stt1939}, \href
  {https://ui.adsabs.harvard.edu/abs/2014MNRAS.437..804J} {437, 804}

\bibitem[\protect\citeauthoryear{{Joshi}, {Dambis}, {Pandey}  \&
  {Joshi}}{{Joshi} et~al.}{2016}]{2016A&A...593A.116J}
{Joshi} Y.~C.,  {Dambis} A.~K.,  {Pandey} A.~K.,   {Joshi} S.,  2016, \mn@doi
  [\aap] {10.1051/0004-6361/201628944}, \href
  {https://ui.adsabs.harvard.edu/abs/2016A&A...593A.116J} {593, A116}

\bibitem[\protect\citeauthoryear{{Kharchenko}, {Piskunov}, {R{\"o}ser},
  {Schilbach}  \& {Scholz}}{{Kharchenko} et~al.}{2005}]{2005A&A...438.1163K}
{Kharchenko} N.~V.,  {Piskunov} A.~E.,  {R{\"o}ser} S.,  {Schilbach} E.,
  {Scholz} R.~D.,  2005, \mn@doi [\aap] {10.1051/0004-6361:20042523}, \href
  {https://ui.adsabs.harvard.edu/abs/2005A&A...438.1163K} {438, 1163}

\bibitem[\protect\citeauthoryear{{Kharchenko}, {Piskunov}, {Schilbach},
  {R{\"o}ser}  \& {Scholz}}{{Kharchenko} et~al.}{2013}]{2013A&A...558A..53K}
{Kharchenko} N.~V.,  {Piskunov} A.~E.,  {Schilbach} E.,  {R{\"o}ser} S.,
  {Scholz} R.~D.,  2013, \mn@doi [\aap] {10.1051/0004-6361/201322302}, \href
  {https://ui.adsabs.harvard.edu/abs/2013A&A...558A..53K} {558, A53}

\bibitem[\protect\citeauthoryear{{Kim}, {Figer}, {Lee}  \& {Morris}}{{Kim}
  et~al.}{2000}]{2000ApJ...545..301K}
{Kim} S.~S.,  {Figer} D.~F.,  {Lee} H.~M.,   {Morris} M.,  2000, \mn@doi [\apj]
  {10.1086/317807}, \href
  {https://ui.adsabs.harvard.edu/abs/2000ApJ...545..301K} {545, 301}

\bibitem[\protect\citeauthoryear{{Kuhn}, {Hillenbrand}, {Sills}, {Feigelson}
  \& {Getman}}{{Kuhn} et~al.}{2019}]{2019ApJ...870...32K}
{Kuhn} M.~A.,  {Hillenbrand} L.~A.,  {Sills} A.,  {Feigelson} E.~D.,   {Getman}
  K.~V.,  2019, \mn@doi [\apj] {10.3847/1538-4357/aaef8c}, \href
  {https://ui.adsabs.harvard.edu/abs/2019ApJ...870...32K} {870, 32}

\bibitem[\protect\citeauthoryear{{Kumar}, {Sagar}  \& {Melnick}}{{Kumar}
  et~al.}{2008}]{2008MNRAS.386.1380K}
{Kumar} B.,  {Sagar} R.,   {Melnick} J.,  2008, \mn@doi [\mnras]
  {10.1111/j.1365-2966.2008.12926.x}, \href
  {https://ui.adsabs.harvard.edu/abs/2008MNRAS.386.1380K} {386, 1380}

\bibitem[\protect\citeauthoryear{{Lada} \& {Lada}}{{Lada} \&
  {Lada}}{2003}]{2003ARA&A..41...57L}
{Lada} C.~J.,  {Lada} E.~A.,  2003, \mn@doi [\araa]
  {10.1146/annurev.astro.41.011802.094844}, \href
  {https://ui.adsabs.harvard.edu/abs/2003ARA&A..41...57L} {41, 57}

\bibitem[\protect\citeauthoryear{{Lamers}, {Gieles}  \& {Portegies
  Zwart}}{{Lamers} et~al.}{2005}]{2005A&A...429..173L}
{Lamers} H.~J.~G.~L.~M.,  {Gieles} M.,   {Portegies Zwart} S.~F.,  2005,
  \mn@doi [Astronomy and Astrophysics] {10.1051/0004-6361:20041476}, \href
  {https://ui.adsabs.harvard.edu/abs/2005A&A...429..173L} {429, 173}

\bibitem[\protect\citeauthoryear{{Landolt}}{{Landolt}}{1992}]{1992AJ....104..340L}
{Landolt} A.~U.,  1992, \mn@doi [\aj] {10.1086/116242}, \href
  {https://ui.adsabs.harvard.edu/abs/1992AJ....104..340L} {104, 340}

\bibitem[\protect\citeauthoryear{{Lata}, {Pandey}, {Kesh Yadav}, {Richichi},
  {Irawati}, {Panwar}, {Dhillon}  \& {Marsh}}{{Lata}
  et~al.}{2019}]{2019AJ....158...68L}
{Lata} S.,  {Pandey} A.~K.,  {Kesh Yadav} R.,  {Richichi} A.,  {Irawati} P.,
  {Panwar} N.,  {Dhillon} V.~S.,   {Marsh} T.~R.,  2019, \mn@doi [\aj]
  {10.3847/1538-3881/ab298c}, \href
  {https://ui.adsabs.harvard.edu/abs/2019AJ....158...68L} {158, 68}

\bibitem[\protect\citeauthoryear{{Lenz} \& {Breger}}{{Lenz} \&
  {Breger}}{2005}]{2005CoAst.146...53L}
{Lenz} P.,  {Breger} M.,  2005, \mn@doi [Communications in Asteroseismology]
  {10.1553/cia146s53}, \href
  {https://ui.adsabs.harvard.edu/abs/2005CoAst.146...53L} {146, 53}

\bibitem[\protect\citeauthoryear{{Lindegren} et~al.,}{{Lindegren}
  et~al.}{2018}]{2018A&A...616A...2L}
{Lindegren} L.,  et~al., 2018, \mn@doi [\aap] {10.1051/0004-6361/201832727},
  \href {https://ui.adsabs.harvard.edu/abs/2018A&A...616A...2L} {616, A2}

\bibitem[\protect\citeauthoryear{{Lohr}, {Negueruela}, {Tabernero}, {Clark},
  {Lewis}  \& {Roche}}{{Lohr} et~al.}{2018}]{2018MNRAS.478.3825L}
{Lohr} M.~E.,  {Negueruela} I.,  {Tabernero} H.~M.,  {Clark} J.~S.,  {Lewis}
  F.,   {Roche} P.,  2018, \mn@doi [\mnras] {10.1093/mnras/sty1280}, \href
  {https://ui.adsabs.harvard.edu/abs/2018MNRAS.478.3825L} {478, 3825}

\bibitem[\protect\citeauthoryear{{Loktin} \& {Beshenov}}{{Loktin} \&
  {Beshenov}}{2003}]{2003ARep...47....6L}
{Loktin} A.~V.,  {Beshenov} G.~V.,  2003, \mn@doi [Astronomy Reports]
  {10.1134/1.1538491}, \href
  {https://ui.adsabs.harvard.edu/abs/2003ARep...47....6L} {47, 6}

\bibitem[\protect\citeauthoryear{{Lomb}}{{Lomb}}{1976}]{1976Ap&SS..39..447L}
{Lomb} N.~R.,  1976, \mn@doi [\apss] {10.1007/BF00648343}, \href
  {https://ui.adsabs.harvard.edu/abs/1976Ap&SS..39..447L} {39, 447}

\bibitem[\protect\citeauthoryear{{Luhman}}{{Luhman}}{2012}]{2012ARA&A..50...65L}
{Luhman} K.~L.,  2012, \mn@doi [\araa] {10.1146/annurev-astro-081811-125528},
  \href {https://ui.adsabs.harvard.edu/abs/2012ARA&A..50...65L} {50, 65}

\bibitem[\protect\citeauthoryear{{Lynga} \& {Palous}}{{Lynga} \&
  {Palous}}{1987}]{1987A&A...188...35L}
{Lynga} G.,  {Palous} J.,  1987, \aap, \href
  {https://ui.adsabs.harvard.edu/abs/1987A&A...188...35L} {188, 35}

\bibitem[\protect\citeauthoryear{{Marigo} et~al.,}{{Marigo}
  et~al.}{2017}]{2017ApJ...835...77M}
{Marigo} P.,  et~al., 2017, \mn@doi [\apj] {10.3847/1538-4357/835/1/77}, \href
  {https://ui.adsabs.harvard.edu/abs/2017ApJ...835...77M} {835, 77}

\bibitem[\protect\citeauthoryear{{Mart{\'\i}n}, {Lodieu}, {Pavlenko}  \&
  {B{\'e}jar}}{{Mart{\'\i}n} et~al.}{2018}]{2018ApJ...856...40M}
{Mart{\'\i}n} E.~L.,  {Lodieu} N.,  {Pavlenko} Y.,   {B{\'e}jar} V. J.~S.,
  2018, \mn@doi [\apj] {10.3847/1538-4357/aaaeb8}, \href
  {https://ui.adsabs.harvard.edu/abs/2018ApJ...856...40M} {856, 40}

\bibitem[\protect\citeauthoryear{{Mathieu} \& {Latham}}{{Mathieu} \&
  {Latham}}{1986}]{1986AJ.....92.1364M}
{Mathieu} R.~D.,  {Latham} D.~W.,  1986, \mn@doi [\aj] {10.1086/114269}, \href
  {https://ui.adsabs.harvard.edu/abs/1986AJ.....92.1364M} {92, 1364}

\bibitem[\protect\citeauthoryear{{Mayne} \& {Naylor}}{{Mayne} \&
  {Naylor}}{2008}]{2008MNRAS.386..261M}
{Mayne} N.~J.,  {Naylor} T.,  2008, \mn@doi [\mnras]
  {10.1111/j.1365-2966.2008.13025.x}, \href
  {https://ui.adsabs.harvard.edu/abs/2008MNRAS.386..261M} {386, 261}

\bibitem[\protect\citeauthoryear{{Mermilliod}}{{Mermilliod}}{1987}]{1987A&AS...71..413M}
{Mermilliod} J.~C.,  1987, \aaps, \href
  {https://ui.adsabs.harvard.edu/abs/1987A&AS...71..413M} {71, 413}

\bibitem[\protect\citeauthoryear{{Messina}, {Parihar}, {Koo}, {Kim}, {Rey}  \&
  {Lee}}{{Messina} et~al.}{2010}]{2010A&A...513A..29M}
{Messina} S.,  {Parihar} P.,  {Koo} J.~R.,  {Kim} S.~L.,  {Rey} S.~C.,   {Lee}
  C.~U.,  2010, \mn@doi [\aap] {10.1051/0004-6361/200912373}, \href
  {https://ui.adsabs.harvard.edu/abs/2010A&A...513A..29M} {513, A29}

\bibitem[\protect\citeauthoryear{{Michalska}}{{Michalska}}{2019}]{2019MNRAS.487.3505M}
{Michalska} G.,  2019, \mn@doi [\mnras] {10.1093/mnras/stz1500}, \href
  {https://ui.adsabs.harvard.edu/abs/2019MNRAS.487.3505M} {487, 3505}

\bibitem[\protect\citeauthoryear{{Mowlavi}, {Barblan}, {Saesen}  \&
  {Eyer}}{{Mowlavi} et~al.}{2013}]{2013A&A...554A.108M}
{Mowlavi} N.,  {Barblan} F.,  {Saesen} S.,   {Eyer} L.,  2013, \mn@doi [\aap]
  {10.1051/0004-6361/201321065}, \href
  {https://ui.adsabs.harvard.edu/abs/2013A&A...554A.108M} {554, A108}

\bibitem[\protect\citeauthoryear{{Neckel} \& {Chini}}{{Neckel} \&
  {Chini}}{1981}]{1981A&AS...45..451N}
{Neckel} T.,  {Chini} R.,  1981, \aaps, \href
  {https://ui.adsabs.harvard.edu/abs/1981A&AS...45..451N} {45, 451}

\bibitem[\protect\citeauthoryear{{Nilakshi}, {Sagar}, {Pandey}  \&
  {Mohan}}{{Nilakshi} et~al.}{2002}]{2002A&A...383..153N}
{Nilakshi} {Sagar} R.,  {Pandey} A.~K.,   {Mohan} V.,  2002, \mn@doi [\aap]
  {10.1051/0004-6361:20011719}, \href
  {https://ui.adsabs.harvard.edu/abs/2002A&A...383..153N} {383, 153}

\bibitem[\protect\citeauthoryear{{Offner}, {Clark}, {Hennebelle}, {Bastian},
  {Bate}, {Hopkins}, {Moraux}  \& {Whitworth}}{{Offner}
  et~al.}{2014}]{2014prpl.conf...53O}
{Offner} S.~S.~R.,  {Clark} P.~C.,  {Hennebelle} P.,  {Bastian} N.,  {Bate}
  M.~R.,  {Hopkins} P.~F.,  {Moraux} E.,   {Whitworth} A.~P.,  2014, in
  {Beuther} H.,  {Klessen} R.~S.,  {Dullemond} C.~P.,   {Henning} T.,  eds,
  Protostars and Planets VI. p.~53 (\mn@eprint {arXiv} {1312.5326}),
  \mn@doi{10.2458/azu_uapress_9780816531240-ch003}

\bibitem[\protect\citeauthoryear{{Phelps} \& {Janes}}{{Phelps} \&
  {Janes}}{1994}]{1994ApJS...90...31P}
{Phelps} R.~L.,  {Janes} K.~A.,  1994, \mn@doi [\apjs] {10.1086/191857}, \href
  {https://ui.adsabs.harvard.edu/abs/1994ApJS...90...31P} {90, 31}

\bibitem[\protect\citeauthoryear{{Piatti}, {Angelo}  \& {Dias}}{{Piatti}
  et~al.}{2019}]{2019MNRAS.tmp.1972P}
{Piatti} A.~E.,  {Angelo} M.~S.,   {Dias} W.~S.,  2019, \mn@doi [\mnras]
  {10.1093/mnras/stz2050}, \href
  {https://ui.adsabs.harvard.edu/abs/2019MNRAS.tmp.1972P} {p.~1972}

\bibitem[\protect\citeauthoryear{{Piskunov}, {Kharchenko}, {R{\"o}ser},
  {Schilbach}  \& {Scholz}}{{Piskunov} et~al.}{2006}]{2006A&A...445..545P}
{Piskunov} A.~E.,  {Kharchenko} N.~V.,  {R{\"o}ser} S.,  {Schilbach} E.,
  {Scholz} R.~D.,  2006, \mn@doi [\aap] {10.1051/0004-6361:20053764}, \href
  {https://ui.adsabs.harvard.edu/abs/2006A&A...445..545P} {445, 545}

\bibitem[\protect\citeauthoryear{{Piskunov}, {Schilbach}, {Kharchenko},
  {R{\"o}ser}  \& {Scholz}}{{Piskunov} et~al.}{2008}]{2008A&A...477..165P}
{Piskunov} A.~E.,  {Schilbach} E.,  {Kharchenko} N.~V.,  {R{\"o}ser} S.,
  {Scholz} R.~D.,  2008, \mn@doi [\aap] {10.1051/0004-6361:20078525}, \href
  {https://ui.adsabs.harvard.edu/abs/2008A&A...477..165P} {477, 165}

\bibitem[\protect\citeauthoryear{{Rozyczka}, {Thompson}, {Pych}, {Narloch},
  {Poleski}  \& {Schwarzenberg-Czerny}}{{Rozyczka}
  et~al.}{2017}]{2017AcA....67..203R}
{Rozyczka} M.,  {Thompson} I.~B.,  {Pych} W.,  {Narloch} W.,  {Poleski} R.,
  {Schwarzenberg-Czerny} A.,  2017, \mn@doi [\actaa]
  {10.32023/0001-5237/67.3.1}, \href
  {https://ui.adsabs.harvard.edu/abs/2017AcA....67..203R} {67, 203}

\bibitem[\protect\citeauthoryear{{Sagar}, {Piskunov}, {Miakutin}  \&
  {Joshi}}{{Sagar} et~al.}{1986}]{1986MNRAS.220..383S}
{Sagar} R.,  {Piskunov} A.~E.,  {Miakutin} V.~I.,   {Joshi} U.~C.,  1986,
  \mn@doi [\mnras] {10.1093/mnras/220.2.383}, \href
  {https://ui.adsabs.harvard.edu/abs/1986MNRAS.220..383S} {220, 383}

\bibitem[\protect\citeauthoryear{{Sagar}, {Miakutin}, {Piskunov}  \&
  {Dluzhnevskaia}}{{Sagar} et~al.}{1988}]{1988MNRAS.234..831S}
{Sagar} R.,  {Miakutin} V.~I.,  {Piskunov} A.~E.,   {Dluzhnevskaia} O.~B.,
  1988, \mn@doi [\mnras] {10.1093/mnras/234.4.831}, \href
  {https://ui.adsabs.harvard.edu/abs/1988MNRAS.234..831S} {234, 831}

\bibitem[\protect\citeauthoryear{{Sagar}, {Munari}  \& {de Boer}}{{Sagar}
  et~al.}{2001}]{2001MNRAS.327...23S}
{Sagar} R.,  {Munari} U.,   {de Boer} K.~S.,  2001, \mn@doi [\mnras]
  {10.1046/j.1365-8711.2001.04438.x}, \href
  {https://ui.adsabs.harvard.edu/abs/2001MNRAS.327...23S} {327, 23}

\bibitem[\protect\citeauthoryear{{Sahijpal} \& {Kaur}}{{Sahijpal} \&
  {Kaur}}{2018}]{2018MNRAS.481.5350S}
{Sahijpal} S.,  {Kaur} T.,  2018, \mn@doi [\mnras] {10.1093/mnras/sty2612},
  \href {https://ui.adsabs.harvard.edu/abs/2018MNRAS.481.5350S} {481, 5350}

\bibitem[\protect\citeauthoryear{{Salpeter}}{{Salpeter}}{1955}]{1955ApJ...121..161S}
{Salpeter} E.~E.,  1955, \mn@doi [\apj] {10.1086/145971}, \href
  {https://ui.adsabs.harvard.edu/abs/1955ApJ...121..161S} {121, 161}

\bibitem[\protect\citeauthoryear{{Sampedro}, {Dias}, {Alfaro}, {Monteiro}  \&
  {Molino}}{{Sampedro} et~al.}{2017}]{2017MNRAS.470.3937S}
{Sampedro} L.,  {Dias} W.~S.,  {Alfaro} E.~J.,  {Monteiro} H.,   {Molino} A.,
  2017, \mn@doi [\mnras] {10.1093/mnras/stx1485}, \href
  {https://ui.adsabs.harvard.edu/abs/2017MNRAS.470.3937S} {470, 3937}

\bibitem[\protect\citeauthoryear{{Sanner}, {Altmann}, {Brunzendorf}  \&
  {Geffert}}{{Sanner} et~al.}{2000}]{2000A&A...357..471S}
{Sanner} J.,  {Altmann} M.,  {Brunzendorf} J.,   {Geffert} M.,  2000, \aap,
  \href {https://ui.adsabs.harvard.edu/abs/2000A&A...357..471S} {357, 471}

\bibitem[\protect\citeauthoryear{{Scargle}}{{Scargle}}{1982}]{1982ApJ...263..835S}
{Scargle} J.~D.,  1982, \mn@doi [\apj] {10.1086/160554}, \href
  {https://ui.adsabs.harvard.edu/abs/1982ApJ...263..835S} {263, 835}

\bibitem[\protect\citeauthoryear{{Schmidt - Kaler}}{{Schmidt -
  Kaler}}{1982}]{1982ND...26..14S}
{Schmidt - Kaler} T.,  1982, \mn@doi [New series]
  {10.1088/0004-6256/135/5/1934}, \href
  {https://ui.adsabs.harvard.edu/abs/2008AJ....135.1934S} {Group VI, vol. 2b.
  springer - verlag, p. 14}

\bibitem[\protect\citeauthoryear{{Sch{\"o}nrich}, {McMillan}  \&
  {Eyer}}{{Sch{\"o}nrich} et~al.}{2019}]{2019MNRAS.487.3568S}
{Sch{\"o}nrich} R.,  {McMillan} P.,   {Eyer} L.,  2019, \mn@doi [\mnras]
  {10.1093/mnras/stz1451}, \href
  {https://ui.adsabs.harvard.edu/abs/2019MNRAS.487.3568S} {487, 3568}

\bibitem[\protect\citeauthoryear{{Sharma}, {Pandey}, {Ogura}, {Mito},
  {Tarusawa}  \& {Sagar}}{{Sharma} et~al.}{2006}]{2006AJ....132.1669S}
{Sharma} S.,  {Pandey} A.~K.,  {Ogura} K.,  {Mito} H.,  {Tarusawa} K.,
  {Sagar} R.,  2006, \mn@doi [\aj] {10.1086/507094}, \href
  {https://ui.adsabs.harvard.edu/abs/2006AJ....132.1669S} {132, 1669}

\bibitem[\protect\citeauthoryear{{Sharma}, {Pandey}, {Ogura}, {Aoki}, {Pandey},
  {Sandhu}  \& {Sagar}}{{Sharma} et~al.}{2008}]{2008AJ....135.1934S}
{Sharma} S.,  {Pandey} A.~K.,  {Ogura} K.,  {Aoki} T.,  {Pandey} K.,  {Sandhu}
  T.~S.,   {Sagar} R.,  2008, \mn@doi [\aj] {10.1088/0004-6256/135/5/1934},
  \href {https://ui.adsabs.harvard.edu/abs/2008AJ....135.1934S} {135, 1934}

\bibitem[\protect\citeauthoryear{{Siegel}, {LaPorte}, {Porterfield}, {Hagen}
  \& {Gronwall}}{{Siegel} et~al.}{2019}]{2019AJ....158...35S}
{Siegel} M.~H.,  {LaPorte} S.~J.,  {Porterfield} B.~L.,  {Hagen} L. M.~Z.,
  {Gronwall} C.~A.,  2019, \mn@doi [\aj] {10.3847/1538-3881/ab21e1}, \href
  {https://ui.adsabs.harvard.edu/abs/2019AJ....158...35S} {158, 35}

\bibitem[\protect\citeauthoryear{{Skrutskie} et~al.,}{{Skrutskie}
  et~al.}{2006}]{2006AJ....131.1163S}
{Skrutskie} M.~F.,  et~al., 2006, \mn@doi [\aj] {10.1086/498708}, \href
  {https://ui.adsabs.harvard.edu/abs/2006AJ....131.1163S} {131, 1163}

\bibitem[\protect\citeauthoryear{{Smith} \& {Jeffries}}{{Smith} \&
  {Jeffries}}{2012}]{2012MNRAS.420.2884S}
{Smith} R.,  {Jeffries} R.~D.,  2012, \mn@doi [\mnras]
  {10.1111/j.1365-2966.2011.20032.x}, \href
  {https://ui.adsabs.harvard.edu/abs/2012MNRAS.420.2884S} {420, 2884}

\bibitem[\protect\citeauthoryear{{Sneden}, {Gehrz}, {Hackwell}, {York}  \&
  {Snow}}{{Sneden} et~al.}{1978}]{1978APJ...223..168S}
{Sneden} C.,  {Gehrz} R.~D.,  {Hackwell} J.~A.,  {York} D.~G.,   {Snow} T.~P.,
  1978, \mn@doi [\apj] {10.1086/156247}, \href
  {https://ui.adsabs.harvard.edu/abs/1978ApJ...223..168S} {223, 168}

\bibitem[\protect\citeauthoryear{{Spitzer}}{{Spitzer}}{1987}]{1987degc.book.....S}
{Spitzer} L.,  1987, {Dynamical evolution of globular clusters}

\bibitem[\protect\citeauthoryear{{Spitzer} \& {Hart}}{{Spitzer} \&
  {Hart}}{1971}]{1971ApJ...164..399S}
{Spitzer} Lyman J.,  {Hart} M.~H.,  1971, \mn@doi [\apj] {10.1086/150855},
  \href {https://ui.adsabs.harvard.edu/abs/1971ApJ...164..399S} {164, 399}

\bibitem[\protect\citeauthoryear{{Stassun} \& {Torres}}{{Stassun} \&
  {Torres}}{2018}]{2018ApJ...862...61S}
{Stassun} K.~G.,  {Torres} G.,  2018, \mn@doi [\apj]
  {10.3847/1538-4357/aacafc}, \href
  {https://ui.adsabs.harvard.edu/abs/2018ApJ...862...61S} {862, 61}

\bibitem[\protect\citeauthoryear{{Stetson}}{{Stetson}}{1992}]{1992ASPC...25..297S}
{Stetson} P.~B.,  1992, in {Worrall} D.~M.,  {Biemesderfer} C.,   {Barnes} J.,
  eds,  Astronomical Society of the Pacific Conference Series Vol. 25,
  Astronomical Data Analysis Software and Systems I. p.~297

\bibitem[\protect\citeauthoryear{{Topasna}, {Kaltcheva}  \&
  {Paunzen}}{{Topasna} et~al.}{2018}]{2018A&A...615A.166T}
{Topasna} G.~A.,  {Kaltcheva} N.~T.,   {Paunzen} E.,  2018, \mn@doi [\aap]
  {10.1051/0004-6361/201731903}, \href
  {https://ui.adsabs.harvard.edu/abs/2018A&A...615A.166T} {615, A166}

\bibitem[\protect\citeauthoryear{{Torres}}{{Torres}}{2010}]{2010AJ....140.1158T}
{Torres} G.,  2010, \mn@doi [\aj] {10.1088/0004-6256/140/5/1158}, \href
  {https://ui.adsabs.harvard.edu/abs/2010AJ....140.1158T} {140, 1158}

\bibitem[\protect\citeauthoryear{{Venuti} et~al.,}{{Venuti}
  et~al.}{2015}]{2015A&A...581A..66V}
{Venuti} L.,  et~al., 2015, \mn@doi [\aap] {10.1051/0004-6361/201526164}, \href
  {https://ui.adsabs.harvard.edu/abs/2015A&A...581A..66V} {581, A66}

\bibitem[\protect\citeauthoryear{{Watson}, {Henden}  \& {Price}}{{Watson}
  et~al.}{2017}]{2017yCat....102027W}
{Watson} C.,  {Henden} A.~A.,   {Price} A.,  2017, VizieR Online Data Catalog,
  \href {https://ui.adsabs.harvard.edu/abs/2017yCat....102027W} {p. B/vsx}

\bibitem[\protect\citeauthoryear{{Wu}, {Zhou}, {Ma}  \& {Du}}{{Wu}
  et~al.}{2009}]{2009MNRAS.399.2146W}
{Wu} Z.-Y.,  {Zhou} X.,  {Ma} J.,   {Du} C.-H.,  2009, \mn@doi [\mnras]
  {10.1111/j.1365-2966.2009.15416.x}, \href
  {https://ui.adsabs.harvard.edu/abs/2009MNRAS.399.2146W} {399, 2146}

\bibitem[\protect\citeauthoryear{{Yadav}, {Sariya}  \& {Sagar}}{{Yadav}
  et~al.}{2013}]{2013MNRAS.430.3350Y}
{Yadav} R.~K.~S.,  {Sariya} D.~P.,   {Sagar} R.,  2013, \mn@doi [\mnras]
  {10.1093/mnras/stt136}, \href
  {https://ui.adsabs.harvard.edu/abs/2013MNRAS.430.3350Y} {430, 3350}

\bibitem[\protect\citeauthoryear{{Zejda}, {Paunzen}, {Baumann},
  {Mikul{\'a}{\v{s}}ek}  \& {Li{\v{s}}ka}}{{Zejda}
  et~al.}{2012}]{2012A&A...548A..97Z}
{Zejda} M.,  {Paunzen} E.,  {Baumann} B.,  {Mikul{\'a}{\v{s}}ek} Z.,
  {Li{\v{s}}ka} J.,  2012, \mn@doi [\aap] {10.1051/0004-6361/201219186}, \href
  {https://ui.adsabs.harvard.edu/abs/2012A&A...548A..97Z} {548, A97}

\bibitem[\protect\citeauthoryear{{Zinn}, {Pinsonneault}, {Huber}  \&
  {Stello}}{{Zinn} et~al.}{2019}]{2019ApJ...878..136Z}
{Zinn} J.~C.,  {Pinsonneault} M.~H.,  {Huber} D.,   {Stello} D.,  2019, \mn@doi
  [\apj] {10.3847/1538-4357/ab1f66}, \href
  {https://ui.adsabs.harvard.edu/abs/2019ApJ...878..136Z} {878, 136}

\bibitem[\protect\citeauthoryear{{von Hoerner}}{{von
  Hoerner}}{1957}]{1957ApJ...125..451V}
{von Hoerner} S.,  1957, \mn@doi [\apj] {10.1086/146321}, \href
  {https://ui.adsabs.harvard.edu/abs/1957ApJ...125..451V} {125, 451}

\makeatother
\end{thebibliography}

\end{document}